\documentclass[journal,twoside]{IEEEtran}
\usepackage{amsfonts}

\usepackage{verbatim}
\usepackage[warnundef]{jabbrv}
\usepackage{epsfig}
\usepackage{times}
\usepackage{amsmath}
\usepackage{amssymb}
\usepackage{algorithm,algorithmic}
\usepackage{graphicx}
\usepackage{adjustbox}
\usepackage{epstopdf}
\usepackage{booktabs}
\usepackage{array}
\usepackage{multirow}
\usepackage[T1]{fontenc}
\usepackage[svgnames,table]{xcolor}
\usepackage{url}
\usepackage[hidelinks]{hyperref}
\usepackage{cite}
\usepackage{etoolbox}
\usepackage{float}
\usepackage[caption = false]{subfig}

\DefineJournalAbbreviation{Test}{Tes} 
\DefineJournalAbbreviation{Transactions}{Trans} 
\DefineJournalAbbreviation{Conference}{Conf} 
\DefineJournalAbbreviation{Aerospace}{Aerosp} 
\DeclareMathOperator{\diag}{diag}

\DeclareMathOperator{\vmin}{vmin}
\DeclareMathOperator*{\argmax}{arg\,max}
\newcolumntype{P}[1]{>{\centering\arraybackslash}p{#1}}
\newcolumntype{M}[1]{>{\centering\arraybackslash}m{#1}}
\newcolumntype{L}[1]{>{\raggedright\let\newline\\\arraybackslash\hspace{0pt}}m{#1}}
\newcolumntype{C}[1]{>{\centering\let\newline\\\arraybackslash\hspace{0pt}}m{#1}}
\newcolumntype{R}[1]{>{\raggedleft\let\newline\\\arraybackslash\hspace{0pt}}m{#1}}
\renewcommand{\arraystretch}{1.1}

\newcommand{\ie}{\textit{i.e.}}
\newcommand{\eg}{\textit{e.g.}}

\definecolor{T}{RGB}{0,0,150}
\newcommand{\hlnew}[1]{\textcolor{blue}{#1}} 

\usepackage{amsthm} 

\ifCLASSINFOpdf
\else
\fi

\makeatletter
\patchcmd\@combinedblfloats{\box\@outputbox}{\unvbox\@outputbox}{}{   \errmessage{\noexpand\@combinedblfloats could not be patched}}
 \makeatother

\input{tcilatex}

\begin{document}

\vspace*{\fill}

\begin{minipage}{1.0\textwidth}
	$\copyright$ 2019 IEEE.  Personal use of this material is permitted.  Permission from IEEE must be obtained for all other uses, in any current or future media, including reprinting/republishing this material for advertising or promotional purposes, creating new collective works, for resale or redistribution to servers or lists, or reuse of any copyrighted component of this work in other works.
\end{minipage}

\vspace*{\fill}
\thispagestyle{empty}
\newpage

\title{Online UAV Path Planning for Joint Detection and Tracking of Multiple Radio-tagged Objects}
\author{Hoa Van Nguyen, S. Hamid Rezatofighi, Ba-Ngu Vo, and Damith C.
Ranasinghe 
\thanks{Acknowledgement: This  work  is  supported  by the  Australian  Research  Council  under Linkage Project LP160101177 and Discovery Project DP160104662.}
\thanks{%
Hoa Van Nguyen, S. Hamid Rezatofighi and Damith C. Ranasinghe are with the
School of Computer Science, The University of Adelaide, SA 5005, Australia
(e-mail: {hoavan.nguyen, hamid.rezatofighi,
damith.ranasinghe}@adelaide.edu.au).}\thanks{%
Ba-Ngu Vo is with the Department of Electrical and Computer Engineering,
Curtin University, Bentley, WA 6102, Australia (e-mail:
ba-ngu.vo@curtin.edu.au).} }
\maketitle
\setcounter{page}{1}
\begin{abstract}

We consider the problem of online path planning for joint detection and tracking of multiple unknown radio-tagged objects. This is a necessary task for gathering spatio-temporal information using UAVs with on-board sensors in a range of monitoring applications. In this paper, we propose an online path planning algorithm with joint detection and tracking because signal measurements from these objects are inherently noisy. 
We derive a partially observable Markov decision process with a random finite set track-before-detect (TBD) multi-object filter, which also maintains a safe distance between the UAV and the objects of interest using a void probability constraint. We show that, in practice, the multi-object likelihood function of raw signals received by the UAV in the time-frequency domain is separable and results in a numerically efficient multi-object TBD filter. We derive a TBD filter with a jump Markov system to accommodate maneuvering objects capable of switching between different dynamic modes. Our evaluations demonstrate the capability of the proposed approach to handle multiple radio-tagged objects subject to birth, death, and motion modes. Moreover, this online planning method with the TBD-based filter outperforms its detection-based counterparts in detection and tracking, especially in low signal-to-noise ratio environments.

\end{abstract}


%





\begin{IEEEkeywords}
	POMDP, track-before-detect, received signal strength, information divergence, RFS, UAV.
\end{IEEEkeywords}

%
\IEEEpeerreviewmaketitle


\section{Introduction}

\IEEEPARstart{A}{rguably}, one of the emerging disruptive technologies of
the 21st century is what the Harvard Business Review \cite{Anderson17} has
recently coined the \textquotedblleft Internet of Flying
Things\textquotedblright , referring to the latest generation of consumer
grade drones or UAVs, capable of carrying imaging, thermal or even
chemical/radiation/biological sensors. Drones are touted to be
transformational for tasks from wildlife monitoring, agricultural
inspection, building inspection, to threat detection, as they have the
potential to dramatically reduce both the time and cost associated with a
traditional manual tasking based on human operators. Realizing this
potential requires equipping UAVs with the ability to carry out missions
autonomously.

In this work, we consider the problem of online path planning for UAV based
localization or tracking of a time-varying number of radio-tagged objects. This
is an important basic problem if UAVs are to be able to autonomously
gather spatial-temporal information about the objects of interest such as
animals in wildlife monitoring \cite{kays2011tracking,thomas2012wildlife,cliff2015online,hoa2017icra},
or safety beacons in search-and-rescue missions \cite{gerasenko2001beacon,Murphy2008}. Signals received by the UAV's on-board radio receiver
are used for the detection and tracking of multiple objects in the region of
interest. However, the radio receiver has a limited range, hence, the UAV---with
limited energy supply---needs to move within range of the mobile objects that
are scattered throughout the region. This is extremely challenging because
neither the exact number nor locations of the objects of interest are
available to the UAV.

Detecting and tracking an unknown and time-varying number of moving objects
in low signal-to-noise ratio (SNR) environments is a challenging problem in
itself. Objects of interest such as wildlife and people tend to switch
between various modes of movements in an unpredictable manner. Constraints
on the transmitters such as cost and battery life mean that signals emitted
from radio-tagged objects have very low power, and become unreliable due to
receiver noise, even when they are within receiving range. 
The traditional approach of detection before tracking incurs information loss, and is not
feasible in such low SNR environments. Reducing information loss introduces
far too many false alarms, while reducing the false alarms increases
misdetections and information loss~\cite{lehmann2012recursive}.

Planning the path for a UAV to effectively detect and track multiple objects
in such environments poses additional challenges. Path planning techniques
for tracking a single object are not applicable. Since there are multiple
moving objects appearing and disappearing in the region, following only
certain objects to localize them accurately means that the UAV is likely to
miss many other objects. The important question is: \textit{which objects should the
UAV follow, and for how long before switching to follow other objects or to
search for new objects?} In addition to detection and tracking, the UAV needs
to maintain a safe distance from the objects without exact knowledge of
their locations. For example, in wildlife monitoring, UAV noise will scare
animals away if they move within a close range. We also need to keep in
mind that the UAV itself has limited power supply as well as computing and
communication resources.

Well-known bio-inspired planning algorithms such as genetic algorithm (GA)
and particle swarm optimization (PSO) \cite{Roberge2013} are computationally
expensive and not suitable for online applications. Markov decision process and partially observable Markov decision process (POMDP) are receiving
increasing attention as online planning algorithms over the last few decades
with techniques such as grid-based MDP \cite{Baek2013}, or POMDP with
nominal belief state optimization \cite{ragi2013uav}. Furthermore, at a
conceptual level, the POMDP framework enables direct generalization to
multiple objects via the use of random finite set (RFS) models \cite%
{mahler2007statistical}. Random finite set can be regarded as a special case of point process when the points are not repeated. For more information related to point process theory, please see [30],\cite{moller2003statistical},\cite{daley2007introduction}. This so-called RFS-POMPD is a POMDP with the
information state being the filtering density of the RFS of objects.

RFS-POMDP provides a natural framework that addresses all the challenges of our online UAV path planning problem. Indeed, RFS-POMDP for
multi-object tracking with various information theoretic reward functions
and task-based reward functions have been proposed in \cite%
{ristic2010sensor, ristic2011anote, hoang2014sensor, hoang2015cauchy,
	beard2015void} and \cite{Reza0,Reza1, Reza2}, respectively. This framework
accommodates path planning for tracking an unknown and time-varying number
of objects in a conceptually intuitive manner. In addition, RFS constructs
such as the void probabilities facilitate the incorporation of a safe distance
between the UAV and objects (whose exact locations are unknown) into the
POMDP \cite{beard2015void}. However, these algorithms require detection to
be performed before tracking and hence not applicable to our problem due to
the low SNR.

In our earlier work~\cite{hoa2017icra}, we presented a path planning solution for tracking one object at a time, in a high SNR environment with a fixed number of objects. This solution, also  based on a detection before tracking formulation, is not applicable to the far more challenging problem of  simultaneously tracking an unknown and time-varying number of objects in low SNR.

In this work, we propose an online path planning algorithm for joint
detection and tracking of multiple objects directly from the received radio
signal in low SNR environments.  
This is accomplished by formulating it as a POMDP with an RFS-based track-before-detect (TBD) multi-object filter. 

TBD methods operate on raw, un-thresholded data \cite{ebenezer2016generalized} and are well-suited for tracking in low SNR environments such as infrared, optical \cite{barniv1985dynamic,tonissen1998maximum,rutten2005recursive,vo2010joint}, and radar \cite{buzzi2005track,buzzi2008track,lehmann2012recursive,dunne2013multiple,papi2015generalized}. However, TBD methods are computationally intensive, and TBD for range-only (received signal strength) tracking has not been developed. One of the main innovations of our solution is to convert the raw signals received by the UAV receiver into time-frequency input measurements for the multi-object TBD filter (using the short time Fourier transform). Such signal representation enables us to derive a separable measurement likelihood function that yields a numerically efficient multi-object TBD filter.

In order to accommodate the time-varying modes of movements of the objects, we use a jump Markov system (JMS) to model their dynamics. Further, to maintain a safe distance from the objects, we impose an object avoidance constraint based on the void probability functional in~\cite{beard2015void} for the planning formulation. 

The paper is organized as follows. Section \ref{sec_background} provides the necessary background: problem statement, RFS and POMDP. Section \ref{sec_problem_formulation} establishes the track-before-detect measurement model, and its implementation to track multiple radio-tagged objects using POMDP under constraints. Section \ref{sec_simulation_experiments} details numerical results and comparisons with detection-based methods. Section \ref{sec_conclusion} reports concluding remarks.
\section{Background} \label{sec_background}
In this work, we consider the problem of online trajectory planning for a UAV to optimally detect and track an unknown and time-varying number of radio-tagged objects.  
Our solution to the problem can be formulated in an RFS-POMDP framework with the multi-object filtering density as the information state. Therefore in the following sections we provide an overview of: i) RFS theory; ii) multi-object filtering using RFS; and iii) the POMDP framework. We start with the problem statement.
\subsection{Problem Statement} \label{sec_problem_statement}
The sensor system under consideration consists of a UAV with antenna elements, and a signal processing module. Following the sensor hardware description in~\cite{hoa2017icra}, we present some of its basic components:
\begin{itemize}
	\item UAVs used are commercial, civilian,  low cost, and small form factor platform with physical constraints on maximum linear and rotation speeds and onboard battery life. 
	
	\item The main payload on a UAV is a directional antenna (\eg, Yagi antenna) to capture radio signals. 
	
	\item The signal processing module is a hardware component embodying a software defined radio capable of receiving and processing multiple radio-tag signals simultaneously. 
\end{itemize}

The objects of interest are equipped with radio transmitters with on-off-keying signaling with low transmit power settings. 
This strategy is commonly used in numerous applications such as very high frequency (VHF) collared tags for wildlife tracking\cite{kays2011tracking,thomas2012wildlife,cliff2015online,hoa2017icra}, or safety beacons for search and rescue missions \cite{gerasenko2001beacon, Murphy2008}. The transmitter design and signaling methods are designed to conserve battery power, reduce the cost of the transmitters, increase the transmitters' lifespan as well as reduce installation and maintenance costs. Such a transmitter usually emits a pulse train of period $T_0$. Within this period, the pulse consists of a truncated sine wave with frequency $f$ over the interval $[\tau,\tau + P_w]$, as illustrated later in Fig.~\ref{fig_received_pulse_stft_illustration}. Low power on-off-keying signals are difficult to detect in noisy environments.    

The objects of interest, \eg, people, wildlife, do not follow very predictable trajectories (such as cars, or planes), and most objects, wildlife, for instance, are afraid of the presence of the UAV in their territories. Consequently, the UAV also needs to maintain a safe distance from objects, although getting close to the objects of interest improves tracking accuracy. Consequently, the received signals from the objects of interest are even harder to detect. 


\subsection{Random Finite Set Models}
For notational consistency, we use lowercase letters (%
\textit{e.g. }$x$) for single-object states; capital letters (\textit{e.g. }$X$) represent the multi-object states; bold letters (\textit{e.g. }$\mathbf{x,X}$) are used for labeled states; blackboard letters (\eg,  $~\mathbb{X}$) denote state spaces. Let $1_A(\cdot)$ denote the inclusion function of a given set $A$, and $\mathcal{F}(A)$ denote the class of finite subset of $A$. If $X = \{ x\}$, for convenience, write $1_{A}(x)$ instead of $1_{A}(\{x\})$. For simplicity, albeit with a slight abuse of notation, we use the symbol $\Phi(\cdot |\cdot )$ to denote the single-object and multi-object transition kernels, and the symbol $g(\cdot |\cdot )$ to denote the single-object and multi-object measurement likelihood functions. 

An RFS $X$ on $\mathbb{X}$ is a random variable taking values in the finite subsets of $\mathbb{X}$. Using Mahler's finite set statistic (FISST), an RFS is fully described by its FISST density. The FISST density is not a probability density \cite{mahler2007statistical}, but it is equivalent to a probability density as shown in \cite{vo2005sequential}. We introduce three common RFS, Bernoulli RFS, multi-Bernoulli RFS and labeled multi-Bernoulli RFS used in our work.
\subsubsection{Bernoulli RFS}
A Bernoulli RFS $X$ on $\mathbb{X}$ has at most one element with probability $r$ for being a singleton distributed over the state space $\mathbb{X}$ according to PDF $p(x)$, and probability $1-r$ for being empty. Its FISST density is defined as follows \cite[p.~351]{mahler2007statistical}: 
\begin{align}
\pi(X) = 
\begin{cases}
1-r & X= \emptyset, \\ \notag
r\cdot p(x) & X = \{x\},%
\end{cases}%
\end{align}
while its cardinality distribution $\rho(n)$ is also a Bernoulli distribution parameterized by $r$.

\subsubsection{Multi-Bernoulli RFS}
is a union of fixed $N$ numbers of independent Bernoulli RFSs: $X =
\bigcup\limits_{i=1}^{N} X^{(i)}$, where $X^{(i)}$ is a random variable on $\mathcal{F}(\mathbb{X}) $ characterized by the existence probability $r^{(i)}$
and probability density $p^{(i)}$ defined on $\mathbb{X}$. Its FISST density is given by~\cite[pp.~368]{mahler2007statistical}:
\begin{align}  
\pi(\{ x^{(1)}, \dots,x^{(n)} \}) &= \pi(\emptyset)
\sum\limits_{1 \leq i_1 \neq \dots \neq i_n \leq N} \prod\limits_{j=1}^{n} \nonumber
\dfrac{r^{(i_{j})} \cdot p^{(i_{j})}(x^{(j)})}{1-r^{(i_{j})}},
\end{align}
where $\pi(\emptyset) = \prod\limits_{i=1}^{N}(1-r^{(i)})$, and its
cardinality distribution is also a multi-Bernoulli distribution ~\cite[pp.~369]{mahler2007statistical}: 
\begin{align}  
\rho(n) = \pi(\emptyset) \sum\limits_{1 \leq i_1 < \dots < i_n \leq N} \notag
\prod\limits_{j=1}^{n} \dfrac{r^{(i_{j})} }{1-r^{(i_{j})}}.
\end{align}

\subsubsection{Labeled Multi-Bernoulli RFS}~\label{sec_LMB_RFS} A labeled RFS with state space $\mathbb{X}$ and label space $\mathbb{L}$ is an RFS on $\mathbb{X} \times \mathbb{L}$ where all realizations of labels are distinct.  Similar to the multi-Bernoulli RFS, a labeled multi-Bernoulli (LMB) RFS is completely defined by a parameter set $\{(r^{(\lambda)},p^{(\lambda)}):\lambda \in \Psi\}$ with index set $\Psi$. Its FISST density is given by: \cite{reuter2014lmb}
\begin{align}
	\mathbf{\pi}(\mathbf{X}) & =\delta_{|\mathbf{X}|}(\mathbf{|\mathcal{L}(\mathbf{X})|})w(\mathcal{L}(\mathbf{X}))p^{\mathbf{X}}, \notag
\end{align}
where 
$\delta$ is the Kronecker delta, $\mathcal{L}(\mathbf{X})$
denotes the set of labels extracted from $\mathbf{X} ~\in \mathcal{F}(\mathbb{X} \times \mathbb{L}) $, $p(\mathbf{x})=p(x,\lambda) = p^{(\lambda)}(x)$,  $p^{\mathbf{X}}=\prod_{(x,\lambda)\in\mathbf{X}}p^{(\lambda)}(x)$,
$w(L)\triangleq\prod_{i\in\mathbb{L}}(1-r^{(i)})\prod_{\lambda\in L}\dfrac{1_{\mathbb{L}}(\lambda)r^{(\lambda)}}{(1-r^{(\lambda)})}$. 
$\\$
\subsection{Multi-object Filtering Using RFS Theory} 
In the FISST approach, the multi-object state at time $k$ is modeled as a (labeled) RFS $\mathbf{X}_k$. The representation of a multi-object state by a finite set
provides consistency with the notion of estimation error distance~\cite{vo2010joint}. Let $z_{1:k}$ denote the history of measurement data from time 1
to $k$. Then using the FISST concept of density and integration, the filtering densities can be propagated using prediction and update steps of the Bayes multi-object filter \cite%
{mahler2007statistical}: 
\begin{align}
& \mathbf{\pi} _{k|k-1}(\mathbf{X}_{k}|z_{1:k-1})  \notag
\label{eq_fisst_bayes_filter_predict} \\
& =\int \mathbf{\Phi}_{k|k-1}(\mathbf{X}_{k}|\mathbf{X}_{k-1})\mathbf{\pi}_{k-1}(%
\mathbf{X}_{k-1}|z_{1:k-1})\delta \mathbf{X}_{k-1}, \\
& \mathbf{\pi} _{k}(\mathbf{X}_{k}|z_{1:k})=\dfrac{g(z_{k}|%
	\mathbf{X}_{k})\mathbf{\pi}_{k|k-1}(\mathbf{X}_{k}|z_{1:k-1})}{\int %
	g(z_{k}|\mathbf{X})\mathbf{\pi}_{k|k-1}(\mathbf{X}|z%
	_{1:k-1})\delta \mathbf{X}}, \label{eq_fisst_bayes_filter_update}
\end{align}%
where $\mathbf{\pi}_{k|k-1}(\cdot|z_{1:k-1})$ denotes a multi-object predicted density; $\mathbf{\pi}_{k}(\cdot|z_{1:k})$ denotes a multi-object filtering density; $\mathbf{\Phi}_{k|k-1}(\cdot |\cdot )$ denotes a transition kernel from time $k-1$ to $k$;  $g(z_{k}|\cdot )$  denotes a measurement likelihood function at time $k$. Note that the multi-object transition kernel $\mathbf{\Phi}_{k|k-1}(\cdot |\cdot )$ incorporates all of dynamic aspects of objects including death, birth and transition to new states. 
The integral is a \textit{set integral} defined for any function $\mathbf{p}:\mathcal{F}(\mathbb{X} \times \mathbb{L}) \rightarrow \mathbb{R}$, given by:
\begin{equation}
\resizebox{0.5\textwidth}{!}{ $\begin{aligned} \label{eq_fisst_definition}
	&\int \mathbf{p(X)} \delta \mathbf{X} =  \notag\\
    &\sum_{n=0}^{\infty}\dfrac{1}{n!} \sum_{(l_1,...,l_n) \in \mathbb{L}^n} \int\limits_{\mathbb{X}^{n}} p(\{(x^{(1)},l^{(1)}), \dots,(x^{(n)},l^{(n)})\}) d(x^{(1)},\dots, x^{(n)})
	\end{aligned}$ }
\end{equation}

Generally, the FISST Bayes multi-object recursion is intractable. However, considerable interest in the field has lead to a number of filtering solutions such as the probability hypothesis density (PHD) filter~\cite{mahler2003phd}, the cardinalized PHD (CPHD) filter~\cite{mahler2007cphd}, the multi-object multi-Bernoulli (MeMBer) filter~\cite{mahler2007statistical,vo2009cardinality}, the
generalized labeled multi-Bernoulli (GLMB) filter~\cite{vo2013glmb,vo2014glmb}, and the labeled multi-Bernoulli (LMB) filter~\cite{reuter2014lmb}.

\subsection{Partially Observable Markov Decision Process} \label{sec_bg_POMDP}
POMDP (partially observable Markov decision process) is a theoretical framework for stochastic control problems and is described by the 6-tuple
$\Big[ \mathcal{F}(\mathbb{T}) \times \mathbb{U}, \mathbb{A}, \mathcal{T},\mathcal{R}, \mathcal{F}(\mathbb{Z}), g(\cdot|\cdot) \Big]$ where \cite{monahan1982state,lovejoy1991survey,bertsekas1996dynamic,Hsu2008}
\begin{itemize}
    \item $\mathbb{T} = \mathbb{X} \times \mathbb{L}$  is the labeled state space;
    \item $\mathcal{F}(\mathbb{T}) \times \mathbb{U}$ is the space where each of its elements is an ordered pair $(\mathbf{X},u)$, with $\mathbf{X}$ is an object state (possibly a multi-object state) and $u$ is an observer state;
    \item $\mathbb{A}$: a set of control actions;
	\item $\mathcal{T}$: a state-transition function on $\big[\mathcal{F}(\mathbb{T}) \times \mathbb{U}\big] \times \mathbb{A} \times \big[ \mathcal{F}(\mathbb{T}) \times \mathbb{U}\big]$ where $\mathcal{T}((\mathbf{X},u),a,(\mathbf{X'},u'))$ is the probability density of next state $(\mathbf{X'},u')$ given current state $(\mathbf{X},u)$ and action taken $a$ by the observer;
	\item $\mathcal{R}$: a real-valued reward function defined on $\mathbb{A}$;
	\item $\mathcal{F}(\mathbb{Z})$: a set of observations;
	\item $g(\cdot|\cdot)$: an observation likelihood function on $\mathcal{F}(\mathbb{Z}) \times \big[\mathcal{F}(\mathbb{T}) \times \mathbb{U}\big] \times \mathbb{A}$ where $g(z|(\mathbf{X},u),a)$ is the likelihood of an observation 
	$z$ given the state $(\mathbf{X},u)$, after the observer takes the action $a$. 
\end{itemize}

The main goal in a POMDP is to find an optimal action $a^{*}_k$ that generates an optimal trajectory (a sequence of observer's positions) by maximizing the total expected reward over $H$ look-ahead steps. Specifically, the total expected reward is $\mathbb{E}[\sum_{j=1}^{H} \gamma^{j-1}\mathcal{R}_{k+j}(a_k)]$  with a discount factor \(\gamma \in (0,1] \) to  modulate the effects of future rewards on current actions, and  $\mathbb{E}[\cdot]$ is the expectation operator.

In this work, we propose using an information-based reward function. For the purpose of joint detection and tracking, where reducing overall uncertainty is the main objective, such a reward function is more appropriate because more information implies less uncertainty~\cite{beard2015void}. There are other reward functions, such as cardinality variances~\cite{Reza0}, which are good at estimating the number of objects or the OSPA-based method in~\cite{Reza1} that depends on user-defined threshold values. In contrast, the information-based methods capture overall cardinality and position information, and can be efficiently computed in a closed-form. A detailed comparison between task-based and information-based reward functions can be found in~\cite{kreucher2005comparison}.

Suppose $\mathbf{\pi}_{k+H|k}(\cdot|z_{1:k})$ is the predicted density to time $k+H$ given measurements up to time $k$, which can be calculated recursively by only using the Bayes prediction step in \eqref{eq_fisst_bayes_filter_predict} from time $k$ to $k+H$. Now, suppose $a_{k}$ is the control action applied to the UAV at time $k$; then, the UAV follows a trajectory consisting  of a sequence of discrete positions $u_{1:H}(a_k) = [u_{k+1}(a_k),\dots, u_{k+H}(a_k)]^T$ with corresponding hypothesized measurements $z_{1:H}(a_k) = [z_{k+1}(a_k),\dots, z_{k+H}(a_k)]^T$.
Then the filtering density $\mathbf{\pi}_{k+H}(\cdot|z_{1:k}, z_{1:H}(a_{k}))$ can be computed recursively using the Bayes filter in \eqref{eq_fisst_bayes_filter_predict}, \eqref{eq_fisst_bayes_filter_update} from time $k$ to $k+H$. The reward function can be specified in terms of information divergence between the filtering density and the predicted density. The rationale is that a more informative filtering density yields better estimation results. Thus, it is appropriate to choose an optimal policy that generates a more informative filtering density. Since the filtering density is equally or more informative than the predicted density, maximizing the information divergence between the filtering density and the predicted density often results in a more informative filtering density, and consequently, a better tracking performance. In particular, the information-based reward function is given by~\cite{beard2015sensor}:
\begin{equation}
\resizebox{0.45\textwidth}{!}{ $\begin{aligned} 
	\mathcal{R}_{k+H}(a_{k}) = D(\mathbf{\pi}_{k+H}(\cdot|z_{1:k},z_{1:H}(a_{k}),\mathbf{\pi}_{k+H|k}(\cdot|z_{1:k})) , \notag
	\end{aligned}$ }
\end{equation}
where  $D(\mathbf{\pi}_2,\mathbf{\pi}_1)$ is the information divergence between two FISST densities, $\mathbf{\pi}_2$ and $\mathbf{\pi}_1$. Some information divergence candidates are R\'{e}nyi divergence (including Kullback-Leibler divergence) or Cauchy-Schwarz divergence described below: 
\subsubsection{R\'{e}nyi divergence} between any two FISST
densities, $\mathbf{\pi}_2$ and $\mathbf{\pi}_1$, is defined as \cite{ristic2010sensor}: 
\begin{align}  
	D_{\text{R\'{e}nyi}}(\mathbf{\pi}_2,\mathbf{\pi}_1) = \dfrac{1}{\alpha-1} \log \int \mathbf{\pi}_2^{\alpha}(\mathbf{X}) \mathbf{\pi}_1^{1-\alpha}(\mathbf{X}) \delta \mathbf{X} \notag
\end{align}
where $\alpha\geq0$ is a parameter  which determines the emphasis of the tails of two distributions in the metric. When $\alpha \rightarrow 1$, we obtain the well known Kullback-Leibler (KL) divergence.
\subsubsection{Cauchy-Schwarz divergence} between any two FISST
densities, $\mathbf{\pi}_2$ and $\mathbf{\pi}_1$, is defined as \cite{hoang2015cauchy}:
\begin{equation} 
\resizebox{0.5\textwidth}{!}{ $\begin{aligned} 
	D_{CS}(\mathbf{\pi}_2,\mathbf{\pi}_1) = -\log \begin{pmatrix}
	\dfrac{\int K^{|\mathbf{X}|} \mathbf{\pi}_2(\mathbf{X}) \mathbf{\pi}_1(\mathbf{X}) \delta \mathbf{X} }{\sqrt{\int K^{|\mathbf{X}|} \mathbf{\pi}_2^2(\mathbf{X}) \delta \mathbf{X} \int K^{|\mathbf{X}|} \pi_1^2(\mathbf{X}) \delta \mathbf{X}}} \notag
	\end{pmatrix}
	\end{aligned}$ }
\end{equation}
where $K$ denotes the unit of hyper-volume on $\mathbb{T}$.

\section{Problem Formulation} \label{sec_problem_formulation}
The problem we formulate involves tracking multiple radio-tagged objects of interest. The state of a single object of interest comprises of all of its kinematic state (denoted as $\zeta=[x,s]^T \in \mathbb{R}^4 \times \mathbb{S} $), including its position and velocity $x \in \mathbb{R}^4$, and its unknown dynamic model $s \in \mathbb{S}$ (\eg, wandering, constant velocity).  Furthermore, each object of interest transmits an on-off-keying signal, as illustrated later in Fig. \ref{fig_received_pulse_stft_illustration}, with unknown offset time $\tau \in \mathbb{R}^+_{0}$ (a non-negative real number), and an unknown unique frequency index $\lambda \in \mathbb{L} \subset \mathbb{N}$ (a natural number). Thus, the state  of a single object of interest is $\mathbf{x} = [\zeta,\tau,\lambda]^T  \in \mathbb{T} =   \mathbb{X} \times \mathbb{L}$, where $\mathbb{X} \subseteq \mathbb{R}^4 \times \mathbb{S} \times \mathbb{R}^+_{0}$.

We begin with a model of the received radio signal in Section~\ref{sec_meas_model} and  derive its separable measurement likelihood function in Section \ref{sec_meas_llm}. We apply our proposed measurement likelihood function to track multiple radio-tagged maneuvering objects in Section \ref{sec_MTT}. We formulate our UAV trajectory planning problem as a POMDP in Section \ref{sec_pomdp_tbd}.

\subsection{Measurement Model} \label{sec_meas_model}
Given a multi-object state $\mathbf{X} \in \mathcal{F}(\mathbb{T})$, each object $\mathbf{x}=[\zeta,\tau,\lambda]^T \in \mathbf{X}$, uniquely identified by frequency index $\lambda$, transmits an on-off-keying signal within a frequency band (\eg, $150-151$ MHz VHF band in Australia for wildlife transmitters \cite{kenward2000manual}) to a directional antenna mounted on an observer. 

\begin{figure}[!tb]
	\centering
	\includegraphics[width=0.5\textwidth]{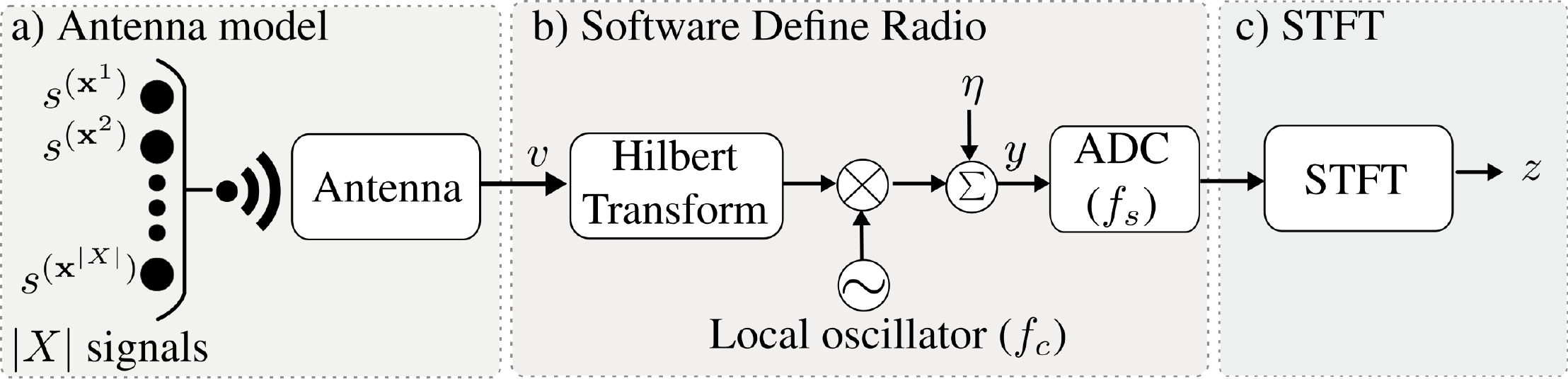}  
	\caption{The receiver model. $|X|$ objects transmit on-off-keying analog signals in time domain. These signals are captured by the antenna and subsequently digitized through a software defined radio device, and converted to time-frequency domain measurements using an STFT algorithm.}
	\label{signal_received_diagram}
\end{figure}

The receiver model of the observer is illustrated in Fig.~\ref{signal_received_diagram}. Here, a Software Defined Radio (SDR) collects received signals from the antenna and down-converts the received signal $v$ via the Hilbert transform and a mixer to a baseband signal $y$, which is subsequently digitized via an embedded analog-to-digital converter (ADC)~\cite{Ossmann2015}. The digitized signal is then transformed to the time-frequency domain via a short time Fourier transform (STFT) algorithm (Fig. \ref{signal_received_diagram}c). In practice, the following holds for the receiver:

\begin{itemize}
	\item The required safety distance between the observer and each object of interest is sufficiently large, so that the transmitted signal can be treated as a far-field signal and the effect of multipath is negligible \cite{hoa2017icra}. 
	\item The receiver noise $\eta$, which may come from the outside environment or thermal noise generated from electronic devices within the receiver, is narrowband wide-sense-stationary (WSS) Gaussian because the bandwidth $B_w$ is small compared to the center frequency $f_c$, $B_w \ll f_c$ \cite[pp.116]{orfanidis2002electromagnetic}. 
\end{itemize}
In the following, we construct a model of the received signals captured by the receiver, and we begin with the antenna model.
$\\$
\textbf{Antenna Model }(Fig. \ref{signal_received_diagram}a): For a single object with state $\mathbf{x}=[\zeta,\tau,\lambda]^T$, the signal $s^{(\mathbf{x})}$  measured at a reference distance $d_0>0$ in the far field region can be modeled as:
\begin{equation} 
\begin{aligned} s^{(\mathbf{x})}(t) = \dfrac{A^{(\lambda)}}{d_0^{\kappa}}\cos[2\pi(f_c + f^{(\lambda)})t +
\phi^{(\lambda)}] \textrm{rect}^{T_0}_{P_w}(t - \tau) \notag, \end{aligned}
\end{equation}
where $A^{(\lambda)}, f^{(\lambda)},\phi^{(\lambda)}$ are the  signal amplitude, baseband frequency and phase, respectively, corresponding to frequency index $\lambda$ of object $\mathbf{x}$; $\kappa $ is a dimensionless path loss exponent that depends on environment and typically ranges from 2 to 4;  $f_c$ is the center frequency of band of interest; where
\begin{equation} \label{eq_rect_def}
\mathrm{rect}^{T_0}_{P_w}(t - \tau) = \sum\limits_{n=-\infty}^{\infty} \mathrm{boxcar}_{\tau}^{\tau+P_w}(t+nT_0)
\end{equation}
is a periodic rectangular pulse train with period $T_0$, pulse width $P_w$; $\mathrm{boxcar}_{a}^{b}(\cdot)$ is a function which is zero over the entire real line except the interval $[a,b]$, where it is equal to unity. 

At the output of directional antenna, the noiseless received signal from a given set $\mathbf{X}$ of objects of interest is modeled as:
\begin{align} \notag
	v^{(u)}(t) = \sum_{\mathbf{x} \in \mathbf{X}}v^{(\mathbf{x},u)}(t).
\end{align}
Here, $v^{(\mathbf{x},u)}$ is the individual signal contribution of object with state $\mathbf{x}$ measured by the observer with state $u$, given by \cite{hoa2017icra}: 
\begin{align}  \label{eq_received_signal_at_antenna_continuous}
&v^{(\mathbf{x},u)}(t) = \\
&\gamma(\zeta,u)\cos[2\pi(f_c + f^{(\lambda)})t +
\psi(\zeta,u)]\mathrm{rect}^{T_0}_{{P_w}}(t -
\tau)\notag,
\end{align}
where
\begin{itemize}
	\item $u = [p^u; \theta^u]$ is the observer state which
	comprises of its position $p^u$ and heading angle $\theta^u$;
	\item $\gamma(\zeta,u) = A^{(\lambda)}G_{r}G_{a}(\zeta,u)(d_{0}/d(p^{\zeta},p^u)
	)^{\kappa}$ is the received signal magnitude when distance between the position of object $\mathbf{x}$ ($p^{\zeta}$) and the position of observer $u$ ($p^u$)  is $d(p^{\zeta},p^u)$;
	\item $G_{r}$ is the receiver gain to amplify the received signal; 
	\item $G_{a}(\zeta,u)$ is the directional antenna gain that depends on a UAV's heading angle $\theta^u$ and its relative position with respect to the position of object $\mathbf{x}$;
	\item $\psi(\zeta,u) = \phi^{(\lambda)} - (f_c +
	f^{(\lambda)}) d(p^{\zeta},p^u)/c$ is the received signal
	phase, where $c$ is the signal velocity.
\end{itemize}
\begin{remark} \label{remark1}\textnormal{
	Notably, the measured signal $v^{(\mathbf{x},u)}$ always depends on the observer state $u$. Hereafter, for notational simplicity, $u$ is suppressed. \eg, $v^{(\mathbf{x})} \triangleq v^{(\mathbf{x},u)}$; $\gamma(\zeta) \triangleq \gamma(\zeta,u) $.  } 
\end{remark}

\noindent\textbf{Software Defined Radio (SDR)} (Fig. \ref{signal_received_diagram}b):
The received signal $v$ is down-converted from the VHF band to the baseband via the Hilbert transform and the mixer. This down-conversion step implemented on the SDR's hardware components is a linear operation and is presented here for completeness. The baseband signal, $\tilde{v}$, is given by: 
\begin{align}
	\tilde{v}(t) = \sum\limits_{\mathbf{x}\in \mathbf{X}} \tilde{v}^{(\mathbf{x})}(t),
\end{align}
where 
\begin{align} \label{eq_tilde_Gamma_i}
	\tilde{v}^{(\mathbf{x})}(t) &\triangleq [v^{(\mathbf{x})}(t) + j [v^{(\mathbf{x})}]^*(t)] e^{-j2\pi f_{c}t} \\
	&= \gamma(\zeta)
	e^{j\psi(\zeta)} e^{j2\pi f^{(\lambda)} t } \mathrm{rect}^{T_0}_{{P_w}}(t -
	\tau); \notag
\end{align}
 $j$ is the imaginary unit; $[v^{(\mathbf{x})}]^*$ is the complex conjugate of $v^{(\mathbf{x})}$. Since the received signal is corrupted by receiver noise $\eta \sim \mathcal{N}(\cdot;0,\Sigma_{\eta})$, the total baseband signal $y$ can be written as:
\begin{align}  \label{eq_meas_time}
y(t) 
&= \sum\limits_{\mathbf{x}\in \mathbf{X}} \tilde{v}^{(\mathbf{x})}(t)+ \eta(t).
\end{align}

This continuous baseband signal $y(\cdot)$ in \eqref{eq_meas_time} is sampled at rate $f_s$ by the ADC component, which generates a discrete-time signal $y[\cdot]$, given by $y[n] \triangleq y(n/f_s)$.
$\\$

\begin{figure}[tb]
	\centering
	\includegraphics[width=0.5\textwidth]{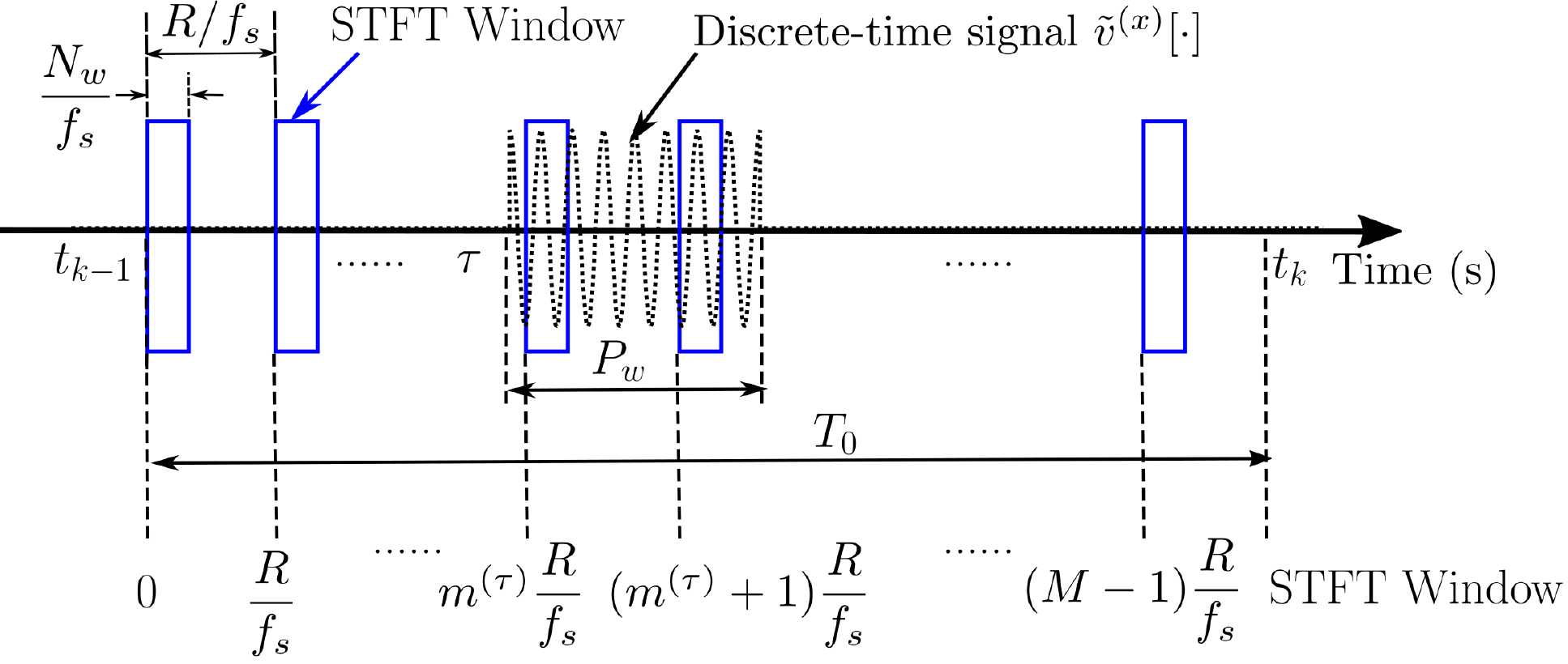} 
	\caption{Illustration for an on-off-keying discrete-time signal $\tilde{v}^{(\mathbf{x})}[\cdot]$ and a STFT windowing method at the $k^\text{th}$ measurement interval  $[t_{k-1}, t_{k})$. $R$ is the hop size, $%
		N_{w}$ is the window width, $P_{w}$ is the pulse width, $\tau$ is the pulse time offset, $T_{0}$ is the period of the pulse. The STFT window frame is indexed at $mR/f_s$ where $m \in \{0,\dots,M-1\}$, and $M$ is the number of window frames in one measurement interval. $m^{(\tau)} = \lceil \tau f_s/R \rceil $ is the time frame index of the signal transmitted from object $\mathbf{x}$. }
	\label{fig_received_pulse_stft_illustration}
\end{figure}

\noindent\textbf{Short-Time Fourier Transform}  (Fig. \ref{signal_received_diagram}c):  The short time Fourier transform (STFT) converts the received signal to a time-frequency measurement. 
Since the on-off keying pulse offset time $\tau$ is unknown, we apply STFT to divide the measurement interval into shorter segments of equal length to capture the sinusoidal component of the received signal to estimate $\tau$ from the measurement. Fig.~\ref{fig_received_pulse_stft_illustration}
illustrates how the STFT is implemented over one measurement interval $[t_{k-1}, t_k)$ of a discrete on-off keying signal (the dash line in Fig. \ref{fig_received_pulse_stft_illustration}) with period $T_0$ and pulse width $P_w$. 
 
To capture the characteristic of the entire signal, we choose the $k^{\text{th}}$ measurement interval to be $[t_k - T_0, t_k)$ to fully contain one cycle of the periodic pulse train. The discrete-time signal on $[t_k - T_0, t_k)$, at the STFT window frame $m \in \{0,\ldots,M-1\}$, is given by: 
\begin{align} \label{eq_discrete_signal}
	y_k^{(m)}[n] \triangleq y(t_k - T_0 + mR/f_s + n/f_s),
\end{align}
where $n = \{0,1,\dots, N_w-1\}$. 

\begin{figure*}[tb]
	\centering
	\includegraphics[width=1.0\textwidth]{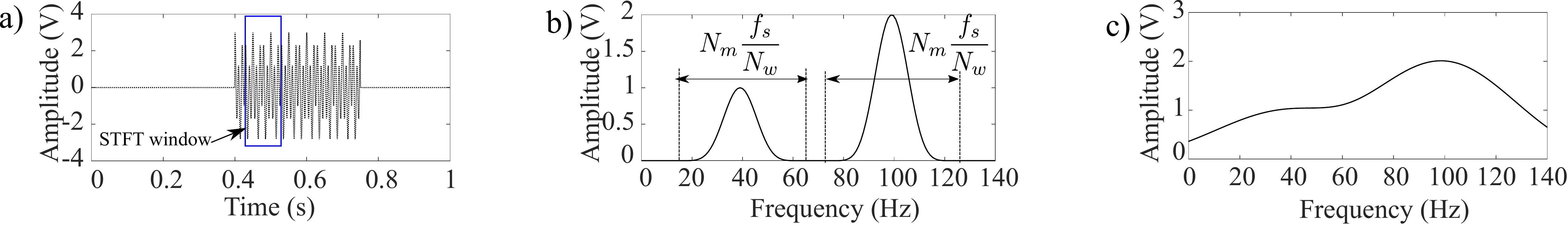}
	\caption{a) Illustration of two on-off-keying signals superpositioned in the time domain: $v(t) = [\cos(40t) + 2\cos(100t)]\mathrm{rect}^{1}_{0.35}(t-0.4)$ at the sampling rate $f_s = 1$~kHz; b) The signals are well-separated in frequency domain when using a 4-term Blackman Harris window with $N_m =8$, where $N_w = 150$ samples and the main-lobe width-in-Hz  $N_m f_s/N_w = 53.33$ Hz < $\triangle f = 60$ Hz; c) However, it is not separable when $N_w = 42$ samples where the main-lobe width-in-Hz $N_m f_s/N_w = 190.47$ Hz > $\triangle f = 60$ Hz.}
	\label{fig_freq_resolution}
\end{figure*}

We set the hop size $R$ and the STFT window width $N_w$ to meet the following condition,
\begin{align} \label{eq_hop_size_Nw_upper_bound}
	1/f^{(\lambda)} \leq N_w < R = P_w f_s /2,
\end{align}
to ensure that the rectangular pulse of signal $\tilde{v}^{(\mathbf{x})}$ in \eqref{eq_tilde_Gamma_i} over the interval $[t_{k}-T_0+\tau,t_{k}-T_0+\tau+P_w)$ contains two non-overlapping STFT window indices, $\{m^{(\tau)},m^{(\tau)}+1\}$ as illustrated in Fig. \ref{fig_received_pulse_stft_illustration}, such that these two STFT windows are only composed of the sinusoidal part of the signal. Thus, the number of window frames in one measurement interval is 
\begin{align} \label{eq_window_frames}
M = \lceil 2T_0/P_w\rceil,
\end{align}
where $\lceil \cdot \rceil$ is the ceiling operator.
The corresponding $L$-point STFT of $y^{(m)}_k[\cdot]$ using the windowing function $w[\cdot]$ is:
\begin{align} \label{eq_meas_DTFT}
	Y^{(m)}_k[l] &= \sum\limits_{n=0}^{N_w-1}
	y^{(m)}_{k}[n]w[n]e^{-j(n+mR)2\pi l/L},
\end{align}
for $l = \{0,1,\dots,L-1\}$ (definitions of different window functions for extracting short-time signal segments and their properties can be found in~\cite{smith2011spectral}).

At the $k^{\text{th}}$ measurement interval, let $\mathbf{X}_k$ denote the multi-object state and $\mathbf{x}_k=[\zeta_k,\tau_k,\lambda_k]^T$ be an element of $\mathbf{X}_k$. By substituting \eqref{eq_rect_def}, \eqref{eq_tilde_Gamma_i}, \eqref{eq_meas_time}, \eqref{eq_discrete_signal} into \eqref{eq_meas_DTFT}, and combining with conditions in \eqref{eq_hop_size_Nw_upper_bound}, $Y^{(m)}_k[l]$ can be written in term of signal and noise components as: 
\begin{align} \label{eq_stft_out_signal_noise}
	Y^{(m)}_k[l] &= \sum_{\mathbf{x}_k \in \mathbf{X}_k} G^{(m,l)}(\mathbf{x}_k) + H^{(m)}_k[l]
\end{align}
where 
\begin{equation} 
\resizebox{0.5\textwidth}{!}{ $\begin{aligned}
	G^{(m,l)}(\mathbf{x}_k)&= 
	\begin{cases}
	\gamma(\zeta_k)e^{j\psi(\zeta_k)}W[l-l^{(\lambda_k)}] & \text{if } m \in \{m^{(\tau_k)},m^{(\tau_k)}+1\} \label{eq_intensity_func} \\ 
	0 & \text{otherwise,}
	\end{cases}\\  \end{aligned} $ }
\end{equation}
\begin{align} 
W[m] &= \sum\limits_{n=0}^{N_w-1} w[n] e^{-jn2\pi l/L}, \label{eq_fft_win_def} \\
l^{(\lambda_k)} &= \lfloor L f^{(\lambda_k)}/f_s \rfloor, \label{eq_freq_idx_def1} \\ 
m^{(\tau_k)} &= \lceil \tau_k f_s/R \rceil, \\
H^{(m)}_k[l] &=\sum\limits_{n=0}^{N_w-1} \eta^{(m)}_{k}[n]w[n] e^{-j(n+mR)2 \pi l/L}, \label{eq_timefreq_noise_frames} \\
\eta^{(m)}_{k}[n] &\triangleq \eta(t_k - T_0 + mR/f_s + n/f_s). \label{eq_stft_of_noise}
\end{align}

Now the measurement data $z_k$ at the $k^{\text{th}}$ measurement interval is an $M\times L$ matrix, with each element $z^{(m,l)}_k = |Y^{(m)}_k[l]|$ is the magnitude of $Y^{(m)}_k[l]$ defined in \eqref{eq_stft_out_signal_noise}.

Notably, to increase the estimation accuracy of the number of transmitted signals, we need to reduce the interference among signal signatures in the frequency domain. Let $N_m$ denote the main-lobe width-in-bins, where each windowing function $w[\cdot]$ affects $N_m$ differently, as shown in Table \ref{table_L_scale_window_type} \cite{smith2011spectral}. Denote $\triangle f$ as the minimum frequency separation among all transmitted signals, given by $\triangle f = \min\limits_{i,j \in \{1,\dots,|\mathbf{X}|\}} |f^{(\lambda^i)} - f^{(\lambda^j)}|$ where $i \neq j$. A criterion to ensure resolvability of signal frequencies requires the main-lobe width-in-Hz of signal signatures be well-separated~\cite{smith2011spectral}, as illustrated in Fig. \ref{fig_freq_resolution}b; hence $N_m f_s/N_w \leq \triangle f$, which implies
\begin{align} \label{eq_Nw_lower_bound}
	 N_w \geq \lceil	N_m \dfrac{f_s}{\triangle f} \rceil.
\end{align}

\begin{table}[!tbp]
	\centering
	\renewcommand{\arraystretch}{1.3}
	\caption{Main-lobe width-in-bins $N_m$ for various windowing functions }
	\label{table_L_scale_window_type}
    \begin{adjustbox}{max width=0.47\textwidth}
	\begin{tabular}{L{1.5cm}|C{1.5cm}C{1.2cm}C{1.2cm}C{2cm}}
		\hline
		\textbf{Windowing Function} & Rectangular & Hamming & Blackman & $B$-term Blackman-Harris \\ \hline
		$N_m$                 & $2$           & $4$       & $6$        & $2B$      \\ \hline
	\end{tabular}
    \end{adjustbox}
\end{table}
Next, we derive the measurement likelihood given measurement $z_k$ and the condition in~\eqref{eq_Nw_lower_bound}.

\subsection{Measurement Likelihood Function} \label{sec_meas_llm}
Let $C(\mathbf{x}_k) $ denote the influence region of an object with state $\mathbf{x}_k$, given by:
\begin{align}
	C(\mathbf{x}_k) \triangleq \{(m,l): |G^{(m,l)}(\mathbf{x}_k)| >0 \}, 
\end{align}
where $G^{(m,l)}(\mathbf{x}_k)$ is defined in \eqref{eq_intensity_func}. We have the following proposition:
\begin{proposition} \label{prop_1}
	\textit{Given a multi-object $\mathbf{X}_k$, and its corresponding measurement $z_k$ at the $k^{\text{th}}$ measurement interval. If the influence region of each object does not overlap, \ie,
	\begin{align}
		C(\mathbf{x}_k)\cap C(\mathbf{x}'_k)=\emptyset ~ \forall~ \mathbf{x}_k,\mathbf{x}'_k \in \mathbf{X}_k,
	\end{align}	
		then the measurement likelihood function is given by: }
	\begin{align} \label{eq_separable_llh}
	g(z_{k}|\mathbf{X}_{k})\propto
	\prod\limits_{\mathbf{x}_k \in \mathbf{X}_k}g_{z_{k}}(\mathbf{x}_{k}),
	\end{align}
	where 
	\begin{itemize}
		\item $g_{z_{k}}(\mathbf{x}_{k})=\prod\limits_{(m,l)\in C(\mathbf{x}_{k})}\dfrac{\varphi (%
			z_{k}^{(m,l)};|G^{(m,l)}(\mathbf{x}_{k})|,\Sigma _{z})}{\phi (%
			z_{k}^{(m,l)};\Sigma _{z})};$
		\item $\varphi(\cdot;|G^{(m,l)}(\mathbf{x}_{k})|,\Sigma_z)$ is the Ricean distribution with mean $|G^{(m,l)}(\mathbf{x}_{k})|$ and covariance $\Sigma_z$;
		\item $\phi(\cdot;\Sigma_z)$ is the Rayleigh distributions with covariance $\Sigma_z$;
		\item $\Sigma_z = E_w \Sigma_{\eta}/2$ is the receiver noise covariance in frequency domain;
		\item  $E_w = \sum\limits_{n=0}^{N_w-1}|w[n]|^2$ is the window energy;
	\end{itemize}
	
\end{proposition}

\indent \textit{Proof:} See the Appendix. 

For our particular problem, given a multi-object $\mathbf{X}$, a single object $\mathbf{x} = [\zeta,~\tau,~\lambda]^T \in \mathbf{X}$ is uniquely identified by the unique frequency index $\lambda$. 
Furthermore, the condition in \eqref{eq_Nw_lower_bound} ensures negligible interference in the frequency domain between the signals emitted from objects with different $\lambda$. As shown in \cite{Harris1978}, using the $4$-term Blackman Harris window, the side-lobe level is less than $-92$ dB compared to the main-lobe level.
Consequently, for all practical purposes, we can consider that the influence region of each object does not overlap, \ie, $ C(\mathbf{x}) \cap C(\mathbf{x}') = \emptyset ~ \forall~\mathbf{x},\mathbf{x}' \in \mathbf{X}$. Thus, Proposition \ref{prop_1} applies to our measurement model. 

\subsection{Multi-object Tracking} \label{sec_MTT}
Tracking an unknown number of objects of interest under noisy measurements is a difficult problem. It is even more challenging when the number of objects of interest may change over time. Due to low power characteristics of signals from radio-tagged objects, the detection-based approaches often fail to detect objects in low signal-to-noise ratio (SNR) environments, especially when objects appear or disappear frequently, which lead to higher tracking errors. Thus, detection based approaches may not be applicable for tracking radio-tagged objects in low SNR environments due to the information loss during the thresholding process to detect objects' signals. On the other hand, the TBD method using raw received signals as measurements, preserves all of the signals' information and has been successfully proven to be an effective filter under low SNR environments in  \cite{barniv1985dynamic,tonissen1998maximum,rutten2005recursive,vo2010joint,buzzi2005track,buzzi2008track,lehmann2012recursive,dunne2013multiple, Papi2013,papi2015generalized}.  

We propose using the TBD-LMB filter~\cite{papi2015generalized} to track a multiple, unknown and time-varying number of objects. 
For our particular problem, the single object state $\mathbf{x} = [\zeta,\tau,\lambda]^T = [\bar{\zeta},\lambda]^T\in \mathbf{X}$ is uniquely identified by $\lambda \in \mathbb{L}$, where $\mathbb{L}$ (assumed to be known)\footnote{In practice, the assumption that $\mathbb{L}$ is known holds; for example, conservation biologists possess a collection of radio-tagged wildlife captured, tagged and released back into the wild. However, $\lambda \in \mathbb{L}$ itself cannot be directly inferred from the measurements, especially under the low signal-to-noise ratio scenarios where existing object signals may or may not be received by the sensor and the sensor also receives interfering measurements (from other users) and thermal noise generated measurement artifacts not originating from any object.}  is a discrete label space containing all frequency indices $\lambda$, and $\bar{\zeta}=[\zeta,\tau]^T \in \mathbb{X}$ is the object state without label.
Hence, the multi-object $\mathbf{X}~\in  \mathcal{F}(\mathbb{T})$ is in fact a labeled RFS. Our initial prior is an LMB density with label space $\mathbb{L}$ and an LMB birth model with label space $\mathbb{B}$ to accommodate an increase in the label space that can occur during UAV path planning for tracking objects\footnote{Notably, in an application where no new objects are introduced into the system over time, the label space $\mathbb{L}$ remains unchanged and the set of LMB birth parameters as expressed in \eqref{eq_MB_predict_params} vanishes. In a practical application, the birth model can accommodate, for example, newly released wildlife during the operation of a tracking task by a UAV.}. Since we use the LMB birth model, TBD-GLMB filter in \cite{papi2015generalized} reduces to a TBD-LMB filter. 


TBD-LMB filter provides a simple and elegant solution for a multi-object tracking approach in a low SNR environment with various tracking uncertainties. However, existing applications of TBD-LMB filters do not make use of a jump Markov system (JMS). 
Following \cite{reuter2015multiple}, we apply a JMS to the proposed TBD-LMB filter by augmenting the discrete mode into the state vector: $\zeta = [x,s]^T $, where $x$ is the object position and velocity, $ s \in \mathbb{S} = \{1,2,...,S_0 \}$ is the object dynamic mode, $S_0 \in \mathbb{N}^+$ is a positive natural number. Moreover, the mode variable is modeled as first-order Markov chain with transitional probability $t_{k|k-1}(s_k|s_{k-1})$. Hence, the
state dynamics and measurement likelihood for a single augmented state vector are given by: 
\begin{align}  \label{eq_jms_tbd_mem_update}
\mathbf{\Phi}_{k|k-1}(\mathbf{x}_{k}|\mathbf{x}_{k-1}) &= 
\Phi_{k|k-1}(\bar{\zeta}_{k}|\bar{\zeta}_{k-1}) \delta_{\lambda_{k-1}}(\lambda_k),  \notag \\
g_{z_{k}}(\mathbf{x}_{k}) &= g_{z_{k}}(x_{k},\tau_k,\lambda_k) = g_{z_{k}}^{(\lambda_k)}(x_{k},\tau_k)\notag,
\end{align}
where 
\begin{align}
\Phi_{k|k-1}(\bar{\zeta}_{k}|\bar{\zeta}_{k-1}) &= \mathcal{N}(x_k;F_{k-1}^{(s_{k-1})}x_{k-1},Q^{(s_{k-1})}) \notag \\
& \times \mathcal{N}(\tau_{k};\tau_{k-1},Q^{(\tau)}) t_{k|k-1}(s_k|s_{k-1}) \nonumber;
\end{align}
$\mathcal{N}(\cdot;\mu,Q)$ denotes a Gaussian density with mean $\mu$ and covariance $Q$; $F_{k-1}^{(s_{k-1})}$ is the single-object dynamic kernel on the discrete mode $s_{k-1}$. The offset time $\tau$ is estimated using a zero mean Gaussian random walk method 
with its covariance $Q^{(\tau)} = \sigma^2_{\tau} {T_0^2}$, where $\sigma^2_{\tau}$ is the standard deviation of the time offset noise. The frequency index $\lambda_k \in \mathbb{L}$ is unique and static, thus the transition kernel for $\lambda_k$ is given by:
\begin{align}
\delta_{\lambda_{k-1}}(\lambda_k) = 
\begin{cases}
1 & \lambda_k = \lambda_{k-1}, \\ 
0 & \text{otherwise}.%
\end{cases}
\end{align} 

\noindent\textbf{LMB Prediction:} At time $k-1$, suppose the filtering density is an LMB RFS described by a parameter set $\mathbf{\pi}_{k-1} = \{r^{(\lambda)}_{k-1},p^{(\lambda)}_{k-1}\}_{\lambda \in \mathbb{L}_{k-1}} $ with state space $\mathbb{X}$ and label space $\mathbb{L}_{k-1}$ (for notational brevity and convenience, $\mathbf{\pi}_{k-1} = \{r^{(\lambda)}_{k-1},p^{(\lambda)}_{k-1}\}_{\lambda \in \mathbb{L}_{k-1}} $ is also used to denote the density of an LMB RFS), and the birth model is also an LMB RFS with a parameter set $\mathbf{\pi}_{B,k} = \{r_{B,k}^{(\lambda)},p_{B,k}^{(\lambda)}\}_{\lambda \in \mathbb{B}_{k}}$ with state space $\mathbb{X}$ and label space $\mathbb{B}_k$, then the predicted multi-object density is also an LMB RFS $\mathbf{\pi}_{k|k-1}  = \{r_{k|k-1}^{(\lambda)},p_{k|k-1}^{(\lambda)}\}_{\lambda \in \mathbb{L}_{k|k-1}}$ with state space $\mathbb{X}$ and label space $\mathbb{L}_{k|k-1} = \mathbb{L}_{k-1} \cup \mathbb{B}_k ~(\text{with } \mathbb{L}_{k-1} \cap \mathbb{B}_k = \emptyset )$, given by the parameter set~\cite{reuter2014lmb}:  
\begin{align}  \label{eq_MB_predict_params}
\mathbf{\pi}_{k|k-1}  &=
\{r_{E,k|k-1}^{(\lambda)},p_{E,k|k-1}^{(\lambda)}\}_{\lambda \in \mathbb{L}_{k-1}}  \cup \{r_{B,k}^{(\lambda)},p_{B,k}^{(\lambda)}\}_{\lambda \in \mathbb{B}_{k}}
\end{align}
where 
\begin{align}  \label{eq_tbd_member_predict_r}
r_{E,k|k-1}^{(\lambda)} =& r_{k-1}^{(\lambda)}\cdot\langle p_{k-1}^{(\lambda)},p^{(\lambda)}_{S,k} \rangle,
\\
p_{E,k|k-1}^{(\lambda)}(\bar{\zeta}) =&  \dfrac{\langle \Phi_{k|k-1}(\bar{\zeta}|\cdot),
p_{k-1}^{(\lambda)}  p^{(\lambda)}_{S,k} \rangle} {\langle p_{k-1}^{(\lambda)},p^{(\lambda)}_{S,k} \rangle}, \label{eq_tbd_member_predict_p}
\end{align}
and $\langle \cdot \rangle$ is the inner product calculated on the previous state $%
\bar{\zeta}_{k-1}$, given by: 
\begin{align}  \label{eq_inner_prod_def}
\langle \alpha, \beta\rangle 
= \sum\limits_{s}  \int  \alpha(x,\tau|s) \beta(x,\tau|s) d(x,\tau). 
\end{align}

\noindent\textbf{LMB Update:} Given the predicted LMB parameters $\mathbf{\pi}_{k|k-1} =
\{r_{k|k-1}^{(\lambda)},p_{k|k-1}^{(\lambda)}\}_{\lambda \in \mathbb{L}_{k|k-1}}$ defined in \eqref{eq_MB_predict_params}, and the measurement likelihood function is separable as in \eqref{eq_separable_llh}, then the filtering  LMB parameters follows~\cite{vo2010joint}: 
\begin{align} \label{eq_MB_update_params}
\mathbf{\pi}_k = \{r_{k}^{(\lambda)},p_{k}^{(\lambda)}\}_{\lambda \in \mathbb{L}_{k|k-1}}	
\end{align}
where
\begin{align}  \label{eq_MB_update_params_r}
r_{k}^{(\lambda)} &= \dfrac{r_{k|k-1}^{(\lambda)}\langle p_{k|k-1}^{(\lambda)}, g_{z_k}^{(\lambda)}\rangle}{1 - r_{k|k-1}^{(\lambda)}+ r_{k|k-1}^{(\lambda)} \langle
p_{k|k-1}^{(\lambda)}, g^{(\lambda)}_{z_k}\rangle}, \\
p_{k}^{(\lambda)} &= \dfrac{ p_{k|k-1}^{(\lambda)} g_{z_k}^{(\lambda)}}{%
\langle p_{k|k-1}^{(\lambda)}, g_{z_k}^{(\lambda)}\rangle}, \label{eq_MB_update_params_p}
\end{align}
with $\langle \cdot \rangle$ is the inner product on the current state $\bar{\zeta}_k$. 

\subsection{Path Planning Under Constraints} \label{sec_pomdp_tbd}

We formulate the online UAV path  planning problem for joint detection and tracking as a partially observable Markov decision process (POMDP) which has been proven
as an efficient and optimal technique for trajectory planning problems~\cite{kaelbling1998planning,castanon2008stochastic}. In the POMDP framework, the purpose of path planning is to find the
optimal policy (\textit{e.g. } a sequence of actions) to maximize the total expected
reward \cite{Reza1}.  Hence, we first focus on evaluating the reward functions. Second, we incorporate a void constraint to maintain a safe distance between the UAV and objects of interest.

\subsubsection{Reward Functions for Path Planning}\label{sec_rewardfunctionsforplaning}

Let $\mathcal{A}_k \in \mathbb{A}$ denote  a set of possible control vectors $a_k$ at time $k$. A common approach is to
calculate an optimal action that maximizes the total expected reward over a look ahead horizon $H$~\cite{ristic2010sensor,hoang2014sensor,beard2015void}---see Section~\ref{sec_bg_POMDP}: 
\begin{align}  \label{eq_reward_argmax}
a^{*}_{k}&=\argmax\limits_{a_{k} \in \mathcal{A}_{k}}
\mathbb{E}\Big[\sum_{j=1}^{H} \gamma^{j-1}\mathcal{R}_{k+j}(a_{k})\Big] 
\end{align}
Since an analytical solution for the expectation of \eqref{eq_reward_argmax} is not available in general, two popular alternatives are to use
 Monte-Carlo integration \cite{beard2015void,ristic2010sensor} or the predicted ideal measurement set (PIMS) as in \cite{ristic2011anote,hoang2014sensor,Amirali2016Cauchy}. Using PIMS, the computationally lower cost approach,  we only generate one ideal future measurement at each measurement interval \cite{hoang2014sensor,Amirali2016Cauchy}. Hence, instead of \eqref{eq_reward_argmax}, the optimal action is defined by:
\begin{align} \label{eq_reward_pims}
	a^{*}_{k}&=\argmax\limits_{a_{k} \in \mathcal{A}_{k}} \sum_{j=1}^{H} \gamma^{j-1} \hat{\mathcal{R}}_{k+j}(a_{k}),
\end{align}
where 
\begin{equation}  
\resizebox{0.42\textwidth}{!}{ $\begin{aligned} \label{eq_reward_sampling_def}
	\hat{\mathcal{R}}_{k+j}(a_{k}) = D(\mathbf{\pi}_{k+j}(\cdot|z_{1:k},\hat{z}_{1:j}(a_{k}), \mathbf{\pi}_{k+j|k}(\cdot|z_{1:k})). 
    \end{aligned} $ }
\end{equation}
In~\eqref{eq_reward_sampling_def}, the predicted density $\mathbf{\pi}_{k+j|k}(\cdot|z_{1:k})$ is calculated by propagating the filtering density $\mathbf{\pi}_k(\cdot|z_{1:k})$ in \eqref{eq_MB_update_params} using the prediction step\footnote{The prediction step generally includes birth, death and object motion. For improving computational time and tractability, we limit this to object motion only as in~\cite{beard2015void}.} in \eqref{eq_tbd_member_predict_r}, \eqref{eq_tbd_member_predict_p} repeatedly, from time $k$ to $k+j$. In contrast, the filtering density $\mathbf{\pi}_{k+j}(\cdot|z_{1:k},\hat{z}_{1:j}(a_{k}))$ is computed recursively by propagating $\mathbf{\pi}_k(\cdot|z_{1:k})$ in \eqref{eq_MB_update_params} from $k$ to $k+j$ using both prediction in \eqref{eq_tbd_member_predict_r}, \eqref{eq_tbd_member_predict_p} and update steps in \eqref{eq_MB_update_params_r}, \eqref{eq_MB_update_params_p} with the ideal measurement $\hat{z}_{1:H}(a_{k})$. The ideal measurement $\hat{z}_{1:j}(a_{k})$ is computed by the following steps \cite{hoang2014sensor}: 
\begin{itemize}
    \item[\textit{i})] Sampling from the filtering density $\mathbf{\pi}_k(\cdot|z_{1:k})$ in \eqref{eq_MB_update_params};
    \item[\textit{ii})] Propagating it to $k+j$ using the prediction step in \eqref{eq_tbd_member_predict_r}, \eqref{eq_tbd_member_predict_p};
    \item[\textit{iii})] Calculating the number of objects $\hat{n}_{k+j|k}$ and the estimated multi-object state $\hat{\mathbf{X}}_{k+j|k} = \{\hat{\mathbf{x}}^{(i)}_{k+j|k}\}_{i=1}^{\hat{n}_{k+j|k}}$;
    \item[\textit{iv})] Simulating the ideal measurement at $k +j$ based on the measurement model in \eqref{eq_stft_out_signal_noise} with the estimated state $\hat{\mathbf{X}}_{k+j|k}$.
\end{itemize}
The number of LMB components for the predicted density $\mathbf{\pi}_{k+j|k}(\cdot|z_{1:k})$ and the filtering density $\mathbf{\pi}_{k+j}(\cdot|z_{1:k},\hat{z}_{1:j}(a_{k}))$ are the same because the measurement likelihood function is separable. For notational simplicity, $\mathbf{\pi}_1 \triangleq \mathbf{\pi}_{k+j|k}(\cdot|z_{1:k})$ and $\mathbf{\pi}_2 \triangleq \mathbf{\pi}_{k+j}(\cdot|z_{1:k},\hat{z}_{1:j}(a_{k}))$ are two LMB densities on $\mathbb{X}$ with the same label space $\mathbb{L}$ (see Section~\ref{sec_LMB_RFS} for a definition of an LMB density), given by:
\begin{align}
\mathbf{\pi}_1 =  \{r_1^{(\lambda)},p_1^{(\lambda)}\}_{\lambda \in \mathbb{L}}; ~ \mathbf{\pi}_2 =  \{r_2^{(\lambda)},p_2^{(\lambda)}\}_{\lambda \in \mathbb{L}};
\end{align}
and rewriting $\mathbf{\pi}_{1}$ and $\mathbf{\pi}_{2}$ in terms of LMB densities:
\begin{align}
\mathbf{\pi}_{1}(\mathbf{X}) & =\delta_{|\mathbf{X}|}(\mathbf{|\mathcal{L}(\mathbf{X})|})w_{1}(\mathcal{L}(\mathbf{X}))p_{1}^{\mathbf{X}}\label{eq:pi_1}\\
\mathbf{\pi}_{2}(\mathbf{X}) & =\delta_{|\mathbf{X}|}(|\mathcal{L}(\mathbf{X})|)w_{2}(\mathcal{L}(\mathbf{X}))p_{2}^{\mathbf{X}}.\label{eq:pi_2}
\end{align}

Hence, evaluating $\hat{\mathcal{R}}_{k+j}(a_{k+j})$ requires calculating the divergence between the two LMB densities $\mathbf{\pi}_{2}$ and $\mathbf{\pi}_{1}$. We consider two candidates to measure  divergence: \textit{i)} R\'{e}nyi divergence; and \textit{ii)} Cauchy-Schwarz divergence described in Section~\ref{sec_bg_POMDP}. However, given the non-linearity of our measurement likelihood, both divergence measures have no closed form solution. Therefore, we approximate the divergence between two LMB densities using Monte-Carlo sampling. In contrast to~\cite{Amirali2016Cauchy} where Monte Carlo sampling was used to approximate the first moment, we approximate the full distribution.

\vspace{2mm}
\noindent1)~\textbf{R\'{e}nyi Divergence Approximation} From the definition in Section~\ref{sec_bg_POMDP}, we have:
\begin{align}
&D_{\text{R\'{e}nyi}}(\mathbf{\pi}_{2},\mathbf{\pi}_{1})  =\dfrac{1}{\alpha-1}\log\int\mathbf{\pi}_{2}^{\alpha}(\mathbf{X})\mathbf{\pi}_{1}^{1-\alpha}(\mathbf{X})\delta\mathbf{X} \notag\\
&=\dfrac{1}{\alpha-1}\log\int\Big[\big(\delta_{|\mathbf{X}|}(\mathbf{|\mathcal{L}(\mathbf{X})|})w_{2}(\mathcal{L}(\mathbf{X}))[p_{2}(\cdot)]^{\mathbf{X}}\big)^{\alpha}  \\
&~~~~~~~~~~~~~~~~~\times 
\big(\delta_{|\mathbf{X}|}(\mathbf{|\mathcal{L}(\mathbf{X})|})w_{1}(\mathcal{L}(\mathbf{X}))[p_{1}(\cdot)]^{\mathbf{X}}\big)^{1-\alpha}\Big]\delta\mathbf{X}. \notag
\end{align}

Since $[p^{\mathbf{X}}]^{\alpha} = \big[\prod_{\mathbf{x}\in \mathbf{X}}p(\mathbf{x})\big]^{\alpha} = \prod_{\mathbf{x}\in \mathbf{X}}[p(\mathbf{x})]^{\alpha} = [p^{\alpha}]^{\mathbf{X}}$, using Lemma 3 in \cite{vo2013glmb}, this becomes:
\begin{align} \label{eq:Renyi_gen}
&D_{\text{R\'{e}nyi}}(\mathbf{\pi}_{2},\mathbf{\pi}_{1})  =\dfrac{1}{\alpha-1}\log\Bigg[\sum_{L\subseteq\mathbb{L}}w_{2}^{\alpha}(L)w_{1}^{1-\alpha}(L) \\
&~~~~~~~~~~~~\times \prod_{\lambda\in L}\Big[\int\big[p_{2}^{(\lambda)}(\bar{\zeta})\big]^{\alpha}\big[p_{1}^{(\lambda)}(\bar{\zeta})\big]^{1-\alpha}d\bar{\zeta}\Big]\Bigg]. \notag
\end{align}

Each $\lambda$ component of $\mathbf{\pi}_{j}$ ($j=1,2$), the continuous density $p_{j}^{(\lambda)}(\cdot)$, is approximated by a probability mass function $\hat{p_{j}}^{(\lambda)}(\cdot)$ using the same set of samples
$\{\bar{\zeta}^{(\lambda,i)}\}_{i=1}^{N_s}$ with different weights $\{\omega_{j}^{(\lambda,i)}\}_{i=1}^{N_s}$:
\begin{equation}
p_{j}^{(\lambda)}(\bar{\zeta})\approx\hat{p}_{j}^{(\lambda)}(\bar{\zeta})=\sum_{i=1}^{N_s}\omega_{j}^{(\lambda,i)}\delta_{\bar{\zeta}^{(\lambda,i)}}(\bar{\zeta}).
\end{equation}
Using Monte Carlo sampling, the product between the two continuous
densities in \eqref{eq:Renyi_gen} can be approximated by the product of two probability mass functions on the finite samples $\{\bar{\zeta}^{(\lambda,i)}\}_{i=1}^{N_s}$, given by: 
\begin{equation}  
\resizebox{0.5\textwidth}{!}{ $\begin{aligned} \label{eq:Renyi_prod}
	&\int\big[p_{2}^{(\lambda)}(\bar{\zeta})\big]^{\alpha}\big[p_{1}^{(\lambda)}(\bar{\zeta})\big]^{1-\alpha}d\bar{\zeta}  \approx\sum_{i=1}^{N_s}\big[\hat{p}{}_{2}^{(\lambda)}(\bar{\zeta}^{(\lambda,i)})\big]^{\alpha}\big[\hat{p}_{1}^{(\lambda)}(\bar{\zeta}^{(\lambda,i)})\big]^{1-\alpha} 
	\\
	&  \approx\sum_{i=1}^{N_s}\big[\sum_{j=1}^{N_s}\omega_{2}^{(\lambda,j)}\delta_{\bar{\zeta}^{(\lambda,j)}}(\bar{\zeta}^{(\lambda,i)})\big]^{\alpha}\big[\sum_{k=1}^{N_s}\omega_{1}^{(\lambda,k)}\delta_{\bar{\zeta}^{(\lambda,k)}}(\bar{\zeta}^{(\lambda,i)})\big]^{1-\alpha}\\
	& \approx\sum_{i=1}^{N_s}\big[\omega_{2}^{(\lambda,i)}\big]^{\alpha}\big[\omega_{1}^{(\lambda,i)}\big]^{1-\alpha},
	\end{aligned} $ }
\end{equation}	
Substituting \eqref{eq:Renyi_prod} into \eqref{eq:Renyi_gen}, the R\'{e}nyi divergence becomes:
\begin{align}
&D_{\textnormal{R\'{e}nyi}}(\mathbf{\pi}_{2},\mathbf{\pi}_{1})\label{eq:Renyi_divergence} \approx\dfrac{1}{\alpha-1} \\
&\times \log\Bigg[\sum_{L\subseteq\mathbb{L}}w_{2}^{\alpha}(L)w_{1}^{1-\alpha}(L)\prod_{\lambda\in L}\Big[\sum_{i=1}^{N_s}\big(\omega_{2}^{(\lambda,i)}\big)^{\alpha}\big(\omega_{1}^{(\lambda,i)}\big)^{1-\alpha}\Big]\Bigg].\notag
\end{align}

\noindent2)~\textbf{Cauchy-Schwarz Divergence Approximation}
From the definition in Section~\ref{sec_bg_POMDP} and following~ \cite{beard2015void}, we have:
\begin{align}
D_{CS}(\mathbf{\pi}_{2},\mathbf{\pi}_{1}) & =-\log\Big(\dfrac{\langle\mathbf{\pi}_{2},\mathbf{\pi}_{1}\rangle_{K}}{\sqrt{\langle\mathbf{\pi}_{2},\mathbf{\pi}_{2}\rangle_{K}\langle\mathbf{\pi}_{1},\mathbf{\pi}_{1}\rangle_{K}}}\Big),\label{eq:cs_divergence}
\end{align}
where 	
\begin{align}
\langle\mathbf{\pi}_{i},\mathbf{\pi}_{j}\rangle_{K} & =\sum_{L\subseteq\mathbb{L}}w_{i}(L)w_{j}(L)\prod_{\lambda\in L}K\langle p_{i}^{(\lambda)}(\cdot),p_{j}^{(\lambda)}(\cdot)\rangle,\label{eq:Cauchy_gen}
\end{align}
for $i,j \in \{1,2\}$. Using the approach in \eqref{eq:Renyi_prod}, we have
\begin{align} \notag
\langle\mathbf{\pi}_{i},\mathbf{\pi}_{j}\rangle_{K} & \approx\sum_{L\subseteq\mathbb{L}}w_{i}(L)w_{j}(L)\prod_{\lambda\in L}K\Big(\sum_{k=1}^{N_s}\omega_{i}^{(\lambda,k)}\omega_{j}^{(\lambda,k)}\Big).\label{eq:cs_divergence_element} 
\end{align}	

The UAV needs to maintain a safe distance from objects, although getting close to the objects of interest improves tracking accuracy. Therefore, in the following section, we derive a void constraint for the path planning formulation.

\subsubsection{Void Probability Functional}

Let $V(u_{k+j}(a_k),r_{\min})$ 
denote the void region of objects 
based on a UAV's position at time $k+j$ if an action $a_k$ is taken. This leads to a cylinder shape where the ground distance between a UAV and any objects should be smaller than $r_{\min}$
, given by:
\begin{equation}
\begin{aligned} 
&V(u_{k+j}(a_k),r_{\min})  = \Big\{ \mathbf{x} \in \mathbb{X}: \label{eq_exclusion_zone_def1} \\ 
&\sqrt{ (p^{(\mathbf{x})}_x - p_x^{(u_{k+j}(a_k))})^2 + (p^{(\mathbf{x})}_y - p_y^{(u_{k+j}(a_k))})^2} < r_{\min} \Big\},  
\end{aligned}
\end{equation}
where $p^{(\mathbf{x})}_x,p^{(\mathbf{x})}_y$ and $p^{(u_{k+j}(a_k))}_x,p^{(u_{k+j}(a_k))}_y$ denote positions of $\mathbf{x}$ and $u_{k+j}(a_k)$ in $x-y$ coordinates, respectively.

Using the closed form expression for the void probability functional\footnote{Here, we use the notion of void probabilities as defined in \cite{kendall1995stochastic}.} of the GLMB in \cite{beard2015void}, we impose the constraint in (44) on the trajectory planning problem as formulated below.


Given a region $S \subseteq \mathbb{X}$ and an LMB density $\mathbf{\pi}$ on $\mathbb{X}$ parameterized as
$\mathbf{\pi} = \delta_{|\mathbf{X}|}(\mathbf{|\mathcal{L}(\mathbf{X})|})w(\mathcal{L}(\mathbf{X}))p^{\mathbf{X}} =  \{r^{(\lambda)},p^{(\lambda)}\}_{\lambda \in \mathbb{L}}$ where each $\lambda$ component is approximated by a set of weighted samples $\{\omega^{(\lambda,i)},\bar{\zeta}^{(\lambda,i)}\}_{i=1}^{N_s}: p^{(\lambda)}(\bar{\zeta}) \approx \sum_{i=1}^{N_s} \omega^{(\lambda,i)} \delta_{\bar{\zeta}^{(\lambda,j)}}(\bar{\zeta})$,  the void functional of $S$ given the multi-object density $\mathbf{\pi}$, $B_{\pi}(S)$, can be approximated as:
\begin{align} \notag 
B_{\mathbf{\pi}}(S) \approx  \sum_{L \subseteq \mathbb{L}} w(\mathcal{L})
\prod_{\lambda \in \mathbb{L}} \big(1-\sum_{i=1}^{N_s}w^{(\lambda,i)} \delta_{\bar{\zeta}^{(\lambda,i)}}(\bar{\zeta}) 1_{S}(\bar{\zeta}) \big)
\end{align}
using the expression of the void probability functional in \cite{beard2015void}. 
Now the maximization problem in \eqref{eq_reward_pims} becomes:
\begin{align}
a^{*}_{k}&=\argmax\limits_{a_{k} \in \mathcal{A}_{k}} 
	\sum_{j=1}^{H} \gamma^{j-1}\hat{\mathcal{R}}_{k+j}(a_{k})
\end{align}
subject to the constraint 
\begin{equation}\notag 
\min\limits_{j\in \{1,\dots ,H\}}[B_{\pi_{k+j}(\cdot|z_{1:k})}(V(u_{k+j}(a_{k}),r_{\min}))]>P_{\vmin}
\end{equation}

\noindent where $P_{\vmin}$ denotes a void probability threshold.

\section{Simulation Experiments} \label{sec_simulation_experiments}

In this section, we evaluate the proposed online path planning strategy for joint detection and tracking of multiple radio-tagged objects using a UAV. 

\subsection{Experimental Settings} \label{sec_dynamic_model}
A two-dimensional area of $[0,1500]~\text{m}\times \lbrack 0,1500]~\text{m}$ is
investigated to demonstrate the proposed approach. The UAV's height is maintained at $30$~m while the objects' heights are fixed at $1$~m to limit the scope to a two-dimensional (2D)
problem\footnote{It can be easily extended to 3D; however, to save computational power, we limit our problem to the 2D domain. }. The total flight time is $400~$s for all experiments. 

\begin{figure*}[!tb]
	\centering
	\includegraphics[width=\textwidth]{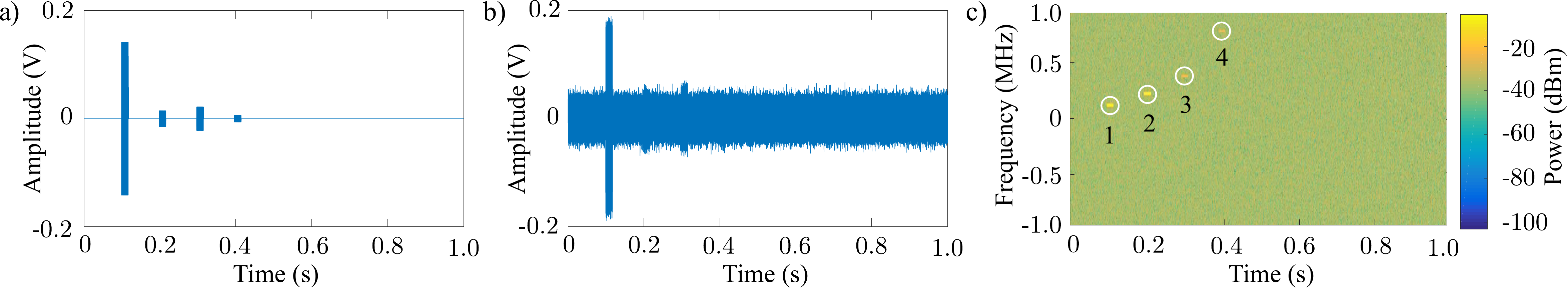}
	\caption{An illustration of received signals from four transmitting objects at distances of $[120,~515,~400,~920]$~m for object 1, object 2, object 3 and object 4 respectively, to the UAV in the presence of complex receiver noise covariance $\Sigma_{\eta} = 0.02^2~V^2$. a) The received signal without noise in time domain; b) the received signal in the presence of the complex white noise in time domain. c) Spectrogram of the received signal in discrete time and frequency domain ($111\times 256$ frames) where the bright spots represent an object's signal in a time-frequency frame.}
	\label{fig_Ex1_Raw_Signal_and_Spectrogram}
\end{figure*}
We also follow the same practical constraints mentioned in \cite{hoa2017icra} for our simulations. The UAV cannot turn its heading instantly, hence its maximum turning rate is limited to $\triangle \theta^u_k = |\theta_k^u -\theta^u_{k-1}| \leq \theta^u_{max}$ (rad/s). In addition, since the planning step normally consumes more time than the tracking step, we apply a cruder planning interval $N_p$ compared to measurement interval $T_0$, such that $N_p = n T_0$ where $n \geq 2,~ n \in \mathbb{N}$ (\ie, $T_0 = 1$ s, $N_p = 5$ s, the planning algorithm calculates the best trajectory for the UAV in next five seconds at each five-measurement-intervals instead of every measurement-interval.)


An object's dynamic mode $s$ follows the jump Markov system where its motion model is either: \textit{i)} a \textit{Wandering} (WD) mode where an object moves short distances without any clear purpose or direction; or \textit{ii)} a constant velocity (CV) mode.

\vspace{2mm}
\noindent\textbf{The Wandering (WD) Model}:  
\begin{align} \label{eq_RW_model}
x_k &= F^{WD}_{k-1}x_{k-1} + q^{WD}_{k-1}
\end{align}
where  $F^{WD}_{k-1} = \diag([1~0~1~0]^T)$, $q^{WD}_{k-1} \sim \mathcal{N}(0,Q^{WD})$ is a zero mean Gaussian process noise with covariance $Q^{WD} =\diag([0.25~\text{m}^2,2.25~\text{(m/s)}^2,~0.25~\text{m}^2,2.25~\text{(m/s)}^2]^T) $. 
In the wandering model, the velocity components are instantly forgotten and then sampled from covariance $Q^{WD}$ at each time step $k$. However, the sampled velocity components do not influence an object's position. Further, the velocity components in $Q^{WD}$ are significantly larger than the position components therein. This is necessary  to achieve the fast moving behavior of objects in the constant velocity dynamic mode when an object switches from the wandering mode to the constant velocity mode.

\vspace{2mm}	
\noindent\textbf{The Constant Velocity (CV) Model}:  
\begin{align} \label{eq_CV_model}
x_k &= F^{CV}_{k-1} x_{k-1} + q^{CV}_{k-1}, \nonumber
\\
F^{CV}_{k-1} &= 
\begin{pmatrix}
1 & T_0 \\ 
0 & 1%
\end{pmatrix}
\otimes I_2,
\end{align}
where $\otimes$ denotes the Kronecker tensor product operator between
two matrices, and $q^{CV}_{k-1} \sim \mathcal{N}(0,Q^{CV})$ is a $%
4\times 1$ zero mean Gaussian process noise, with covariance $Q^{CV} =
\sigma_{CV}^2%
\begin{pmatrix}
T_0^3/3 & T_0^2/2 \\ 
T_0^2/2 & T_0%
\end{pmatrix}%
\otimes I_2$,  where $\sigma_{CV}$ is the standard deviation of
the process noise parameter.

There are four objects with different birth and death times, listed in pairs as $(t_{\rm birth},t_{\rm death})$: $(1,250)$, $(50,300)$, $(100,350)$, $(150,400)$~s. The four objects initially follow the wandering model (WD) with initial state vectors $[800,0.13,300,-1.44]^T$, $[200,0.18,700,-2.17]^T$, $[1200,-1.94,1000,0.42]^T$, $[900,1.91,1300,-2.04]^T$ (with appropriate standard units) at birth. One second period after birth, object 1 and object 3 switch their dynamic mode to the constant velocity mode while object 2 and object 4 continue to follow the wandering model for $65$~s. We detail the mode changes (later) in Fig.~\ref{fig_Ex1_Estimated_Mode}.

For each newly born object, we assume an initial birth state described by a Gaussian distribution with means at $[800,0,300,0]^T$, $[200,0,700,0]^T$, $[1200,0,1000,0]^T$, $[900,0,1300,0]^T$ (with appropriate standard units) and covariance $Q^B=\diag([100~\text{m}^2,4~\text{(m/s)}^2,
100~\text{m}^2,4~\text{(m/s)}^2]^T)$. In practice, such a setting is reasonable and captures the prior knowledge about an object's location. For example, in applications such as wildlife tracking, conservation biologists know the location of newly released wildlife or the locations of entry and exit points of animals that can suddenly appear in a scene from underground animal dwellings. 

All the common parameters used in the following experiments are listed in Tables~\ref{table_birth_death_mode}, \ref{table_signal_parameters}, and \ref{table_measurement_parameter}. 
In addition, Fig. \ref{fig_Ex1_Raw_Signal_and_Spectrogram}a
illustrates a raw received signal without noise from four transmitted objects along with a noisy received signal in Fig. \ref{fig_Ex1_Raw_Signal_and_Spectrogram}b. Furthermore, a single measurement set of the noisy received signal after going through the STFT process consists of $111 \times 256$ time-frequency frames is illustrated in Fig. \ref{fig_Ex1_Raw_Signal_and_Spectrogram}c.

\begin{table}[!tb]
	\centering
	\caption{Birth, death, and dynamic mode parameters}
	\label{table_birth_death_mode}
	\begin{tabular}{L{3.6cm}|L{3.6cm}} \hline
		\textbf{Parameter} & \textbf{Value}  \\ \hline
		Birth probability ($r_B$) & $10^{-6}$ \\ \hline
		Survival probability ($p_S$) & 0.99 \\ \hline
		Initial mode probability & $[0.5~0.5]^T$  \\ \hline
		Mode transitional probability & $[0.99~0.01;0.01~0.99]$ \\ \hline
		Constant velocity noise ($\sigma_{CV}$) & $0.05$ m/$\text{s}^2$ \\ \hline
	\end{tabular}
\end{table}

\begin{table}[!tb]
	\centering
	\caption{Signal parameters}
	\label{table_signal_parameters}
    \begin{adjustbox}{max width=0.45\textwidth}
	\begin{tabular}{L{3cm}|C{1.5cm}|L{3cm}}
		\hline
		\textbf{Parameter} & \textbf{Symbol} & \textbf{Value} \\ \hline
		Center frequency & $f_c$ & $150$ MHz \\ \hline
		Baseband frequencies & $f^{(\lambda)}$ & $131$ kHz, $201$ kHz, $401$kHz , $841$ kHz \\ \hline
		Sampling frequency & $f_s$ & $2$ MHz \\ \hline
		Pulse period & $T_0$ & $1$ s \\ \hline
		Pulse offset time & $\tau^{(\lambda)} $ & $0.1$ s, $0.2$ s, $0.3$ s, $0.4$ s \\ \hline
		Pulse width & $P_w$ & $18$ ms \\ \hline
		Reference distance & $d_0$ & $1$ m \\ \hline
		Pulse amplitude & $A$ & $0.0059$ V \\ \hline
		Path loss constant & $\kappa$ & $3.1068$ \\ \hline
	\end{tabular}
    \end{adjustbox}
\end{table}

\begin{table}[!tb]
	\centering
	\caption{Measurement parameters}
	\label{table_measurement_parameter}
	\begin{tabular}{L{3.5cm}|C{1.5cm}|L{2cm}}
		\hline
		\textbf{Parameter} & \textbf{Symbol} & \textbf{Value} \\ \hline
		Receiver gain & $Gr$ & $72$ dB \\ \hline
		Receiver noise covariance & $\Sigma_{\eta}$ & $0.025^2$ $\text{V}^2$ \\ \hline
		Number of window frames & $M$ & $111$ \\ \hline
		Number of frequency samples & $L$ & $256$ \\ \hline
		Window width & $N_w$ & $256$ \\ \hline
		Number of particles & $N_s$ & $50,000$ \\ \hline
		UAV's max heading angle & $\theta^u_{\max}$ & $\pi/3$ rad/s	 \\ \hline
		UAV's velocity & $v_u$ & $20$ m/s  \\ \hline
		UAV's initial position & $u_1$ & $[0;0;30;\pi/4]$ \\ \hline
		Planning interval & $N_p$ & $5$~s \\ \hline
		Look-a-head horizon & $H$ & 3 \\ \hline
		Minimum distance & $r_{\min}$ & $50$ m \\ \hline
		Void threshold & $P_{\vmin}$ & $0.9$ \\ \hline
		OSPA (order, cut-off) & $(p,c)$ & $(1,100$ m$)$ \\ \hline 
	\end{tabular}
\end{table}

\subsection{Experiments and Results}
We conduct two different experiments: \textit{i)}
to validate and evaluate our proposed planning method for joint detection and tracking; \textit{ii)} compare performance against planning for tracking with conventional detection-then-track methods.
\begin{figure*}[!tb]
	\centering
	\includegraphics[width=0.88\textwidth]{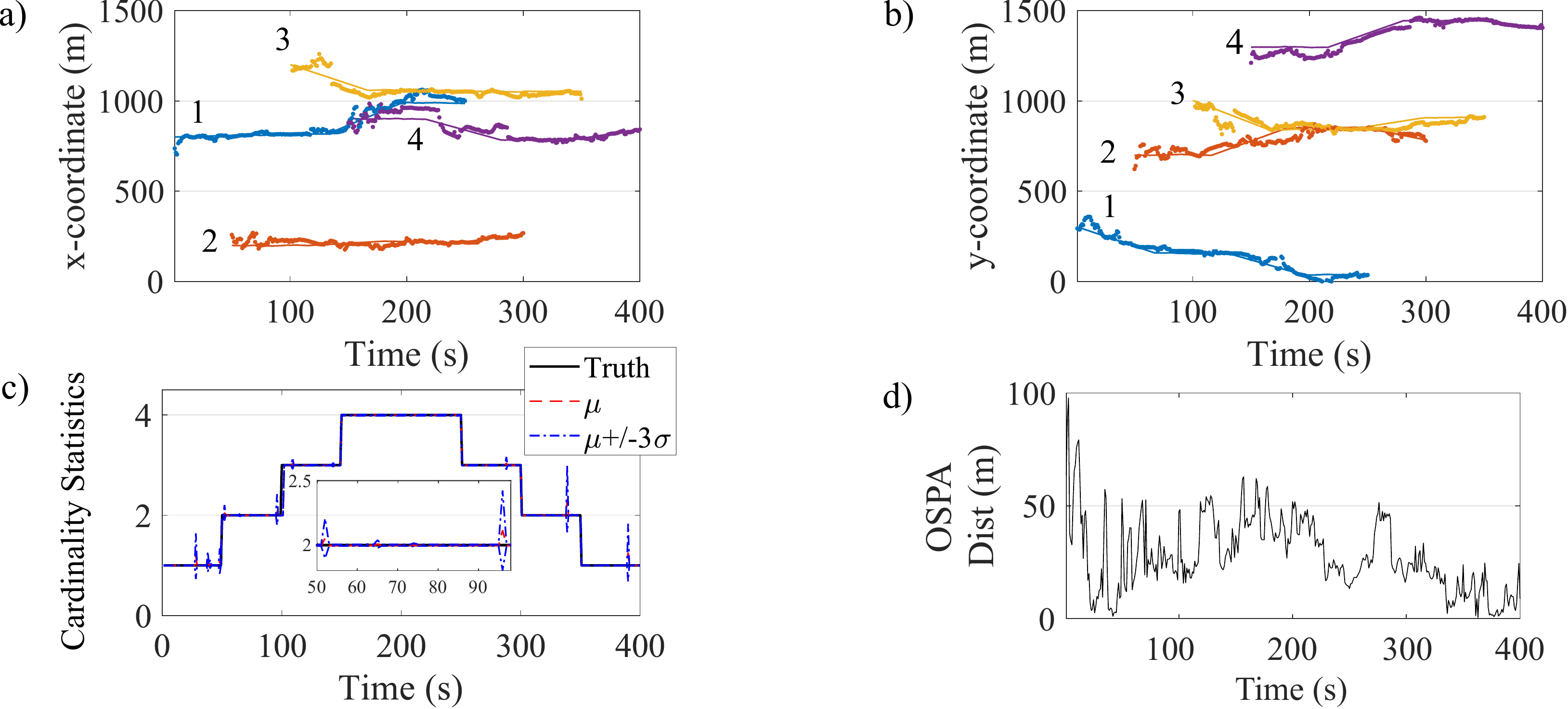}  
	\caption{Tracking four objects in various locations with different birth and death times and motion dynamics. Estimated positions and  truth in: a) x-coordinate; b) y-coordinate;  c) cardinality---its truth versus mean $\mu$ and its variance ($\mu \pm 3\sigma)$; d) OSPA---the cutoff and order parameters are given in Table~\ref{table_measurement_parameter}.}
	\label{fig_Ex1_Estimiated_track_xy_coordinate}
\end{figure*}
\begin{figure}[!ht]
	\centering
	\includegraphics[width=0.5\textwidth]{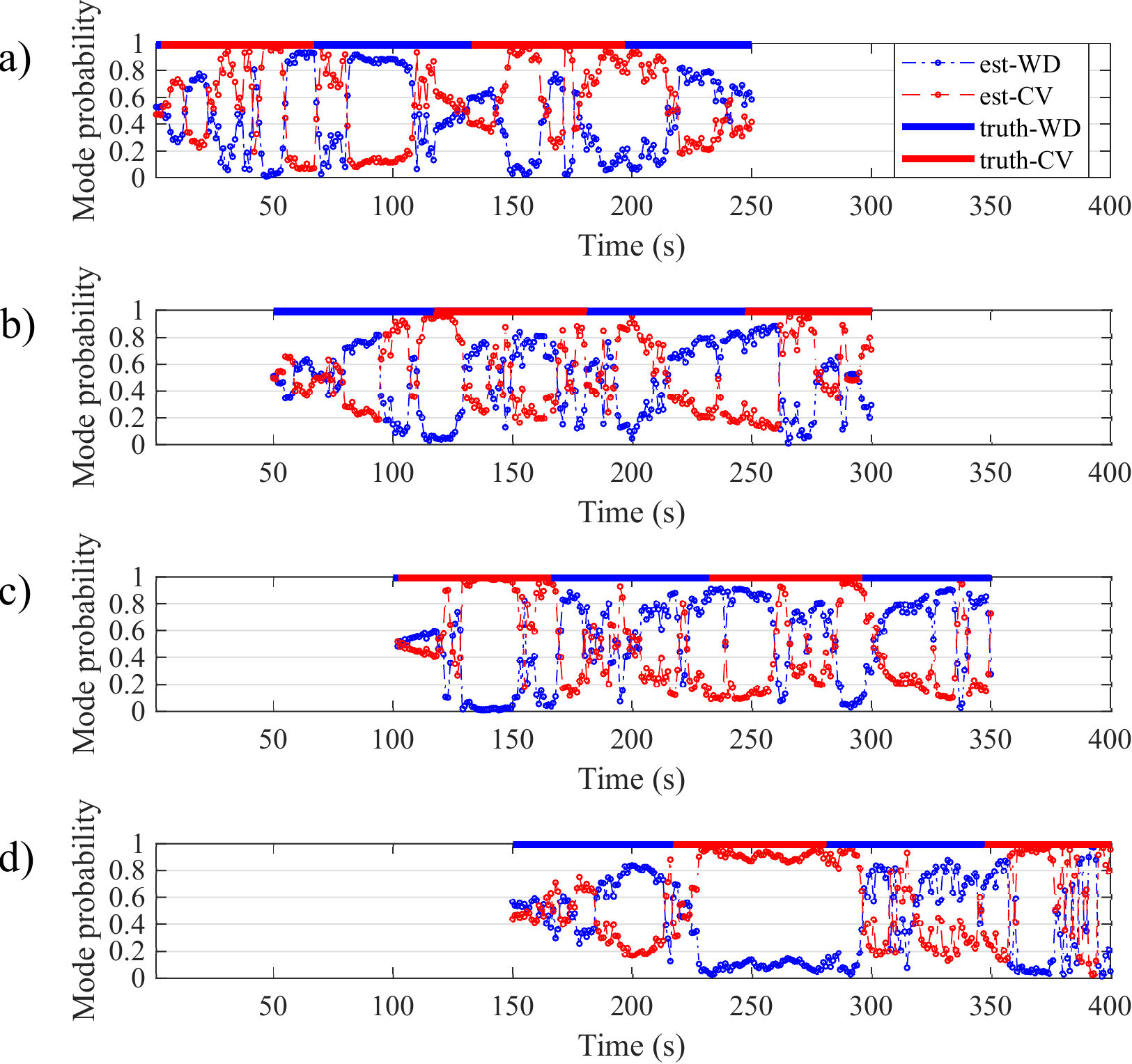}  
    	\caption{The estimated mode probability for four objects with mode WD:Wandering and mode CV:Constant Velocity: a) object $1$; b) object $2$; c) object $3$; and d) object $4$. }
	\label{fig_Ex1_Estimated_Mode}
\end{figure}

\vspace{2mm}
\noindent\textbf{Experiment 1--Validating Planning for Joint Detection and Tracking: } The first experiment is conducted with four objects in various locations and moving in different directions where birth and death times and motion dynamics are described in Section~\ref{sec_dynamic_model}. We employ R\'{e}nyi divergence based reward function with receiver noise covariance $\Sigma_{\eta} = 0.025^2~\text{V}^2$  and the UAV undergoes trajectory changes every 15~s, \ie, the planning interval $N_p = 5$~s with a look ahead horizon $H = 3$ (see Table~\ref{table_measurement_parameter}). 
Fig. \ref{fig_Ex1_Estimiated_track_xy_coordinate}a-b depict true object trajectories, birth and death times together with the estimated tracking accuracy for a typical experiment run. The results show that the proposed planning for joint detection and tracking accurately estimates position and cardinality of the objects. 

Fig.~\ref{fig_Ex1_Estimiated_track_xy_coordinate}c depicts the ground truth changes in the number of objects over time with the estimated cardinality. We used optimal sub-pattern assignment (OSPA)~\cite{schuhmacher2008consistent} to quantify the error between the filter estimates and the ground truth to evaluate the multi-object miss distance. The spikes in Fig.~\ref{fig_Ex1_Estimiated_track_xy_coordinate}c indicate a high uncertainty in the estimated cardinality distribution. The high uncertainty is due to low signal-to-noise ratio (SNR) of received measurements. During path planning, noisy signals lead to poor control decision that result in the UAV navigating to positions further from objects of interest where the signal incident on the UAV sensor antenna is often at an angle where the antenna gain is poor. Further, planning decisions are also subject to void constraints. Consequently, the existence probability of objects of interest can suddenly increase or decrease after a poor control action.

The OSPA distance performance over the tracking period for these objects is depicted in Fig.~\ref{fig_Ex1_Estimiated_track_xy_coordinate}d. We see changes in OSPA distance during birth and death events and its subsequent reduction as the planning algorithm undergoes course changes to improve tracking accuracy.  These results confirm that our trajectory planning algorithm consistently tracks the number of object change over time whilst making course changes to improve estimation accuracy of all the objects.

Fig.~\ref{fig_Ex1_Estimated_Mode} depicts the multiple motion modes of objects and how it changes over time. The results show that although the received signals are noisy, the filter can still accurately estimate the correct mode of objects most of the time.

\begin{figure*}[!tb]
	\centering
	\centering
	\includegraphics[width=\textwidth]{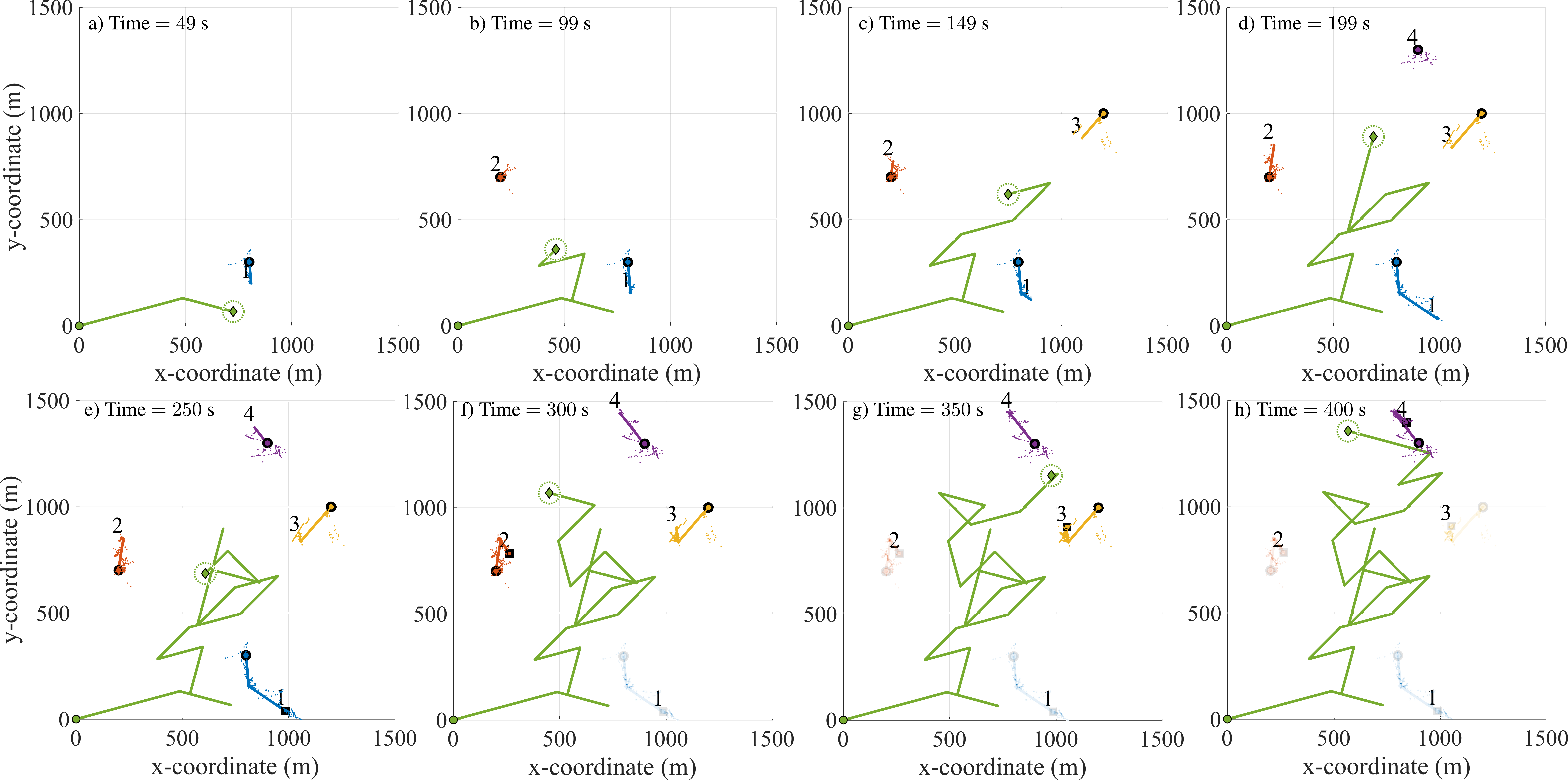}  
	\caption{A typical UAV trajectory (green path) under the proposed path planning for joint detection and tracking algorithm for multiple radio-tagged objects. Here, `$\circ$' locations of object births; `$\Box$' locations of object deaths; `$\Diamond$' current locations of the UAV. Faint tracks show objects subject to a death process.}
	\label{fig_Ex1_Estimated_Positions}
\end{figure*}

Fig.~\ref{fig_Ex1_Estimated_Positions} depicts the evolution of true and estimated object trajectories under the control of the path planning scheme subject to the void constraint. From these snapshots in time, we can see that a typical trajectory to track objects under the birth and death process agrees with our intuition. Initially, the UAV navigates towards object~$1$. At time $t=50$~s (Fig. \ref{fig_Ex1_Estimated_Positions}a), object~$2$ is born; subsequently, the UAV maintains a trajectory between the two objects with course changes to track both objects.  Object~$3$ is born at $t=100$~s, the UAV undertakes course changes to estimate the position of all three moving objects with a maneuver to follow object~$1$ and $2$ whilst moving closer to object $3$ (Fig. \ref{fig_Ex1_Estimated_Positions}b~and~c). We can observe a similar planning strategy evolving when object~$4$ is born at time $150$~s. The UAV navigates to a position to be closer to all four objects and maintain a position at the center of the four objects to estimate the position of all four objects (Fig.~\ref{fig_Ex1_Estimated_Positions}d~and ~e). At time $250$~s, object $1$ vanishes, thus the UAV moves up toward to a position at the center of object $2$, object $3$ and object $4$ to track the remaining objects (Fig.~\ref{fig_Ex1_Estimated_Positions}f). Beyond $300$~s, both object $1$ and object $2$ are no longer in existence; therefore we can observe the UAV heading to a position between objects $3$---whilst maintaining the void constraint illustrated by the dashed circle at the UAV position---and object $4$ (Fig.~\ref{fig_Ex1_Estimated_Positions}g). After time $350$~s, only object $4$ is in existence; thus, the UAV undertakes trajectory changed to move towards object $4$ (Fig. \ref{fig_Ex1_Estimated_Positions}h). The results show that the proposed planning strategy is able to detect and track all objects whilst dynamically acting upon different birth and death events to maneuver the UAV to move to positions that minimize the overall tracking error.

\vspace{2mm}
\noindent\textbf{Experiment 2--Comparing Performance: }
In this experiment, we compare our proposed  online path planning for joint detection and tracking formulation with the TBD-LMB filter with planning for detection-then-track (DTT) methods using a DTT-LMB filter \cite{reuter2014lmb}. We compare three trajectory planning approaches for tracking: \textit{i)} a straight path---direct the UAV back and forth along a diagonal line between $(0,0)$~m and $(1500,1500)$~m; 
\textit{ii)} planning with R\'{e}nyi divergence as the reward function; and \textit{iii)} planning
with Cauchy divergence as the reward function 

The measurements for DTT are extracted based on a peak detection algorithm to find the prominent peak such that the minimum peak separation is $N_m=8$ frequency bins---i.e. the number of main-lobe width-in-bin for a 4-term Blackman Harris window as listed in Table~\ref{table_L_scale_window_type}. Since we examine the filter performance under various receiver noise levels, it is more appropriate to use a peak detection method compared to a fixed threshold value. Further, the peak detection method is robust against different noise levels, considering false alarm and misdetection rates \cite{scholkmann2012efficient}. The planning for DTT methods uses the same PIMS approach for TBD methods listed in Section~\ref{sec_rewardfunctionsforplaning}. 

\begin{figure}[!tb]
	\centering
	\includegraphics[width=0.47\textwidth]{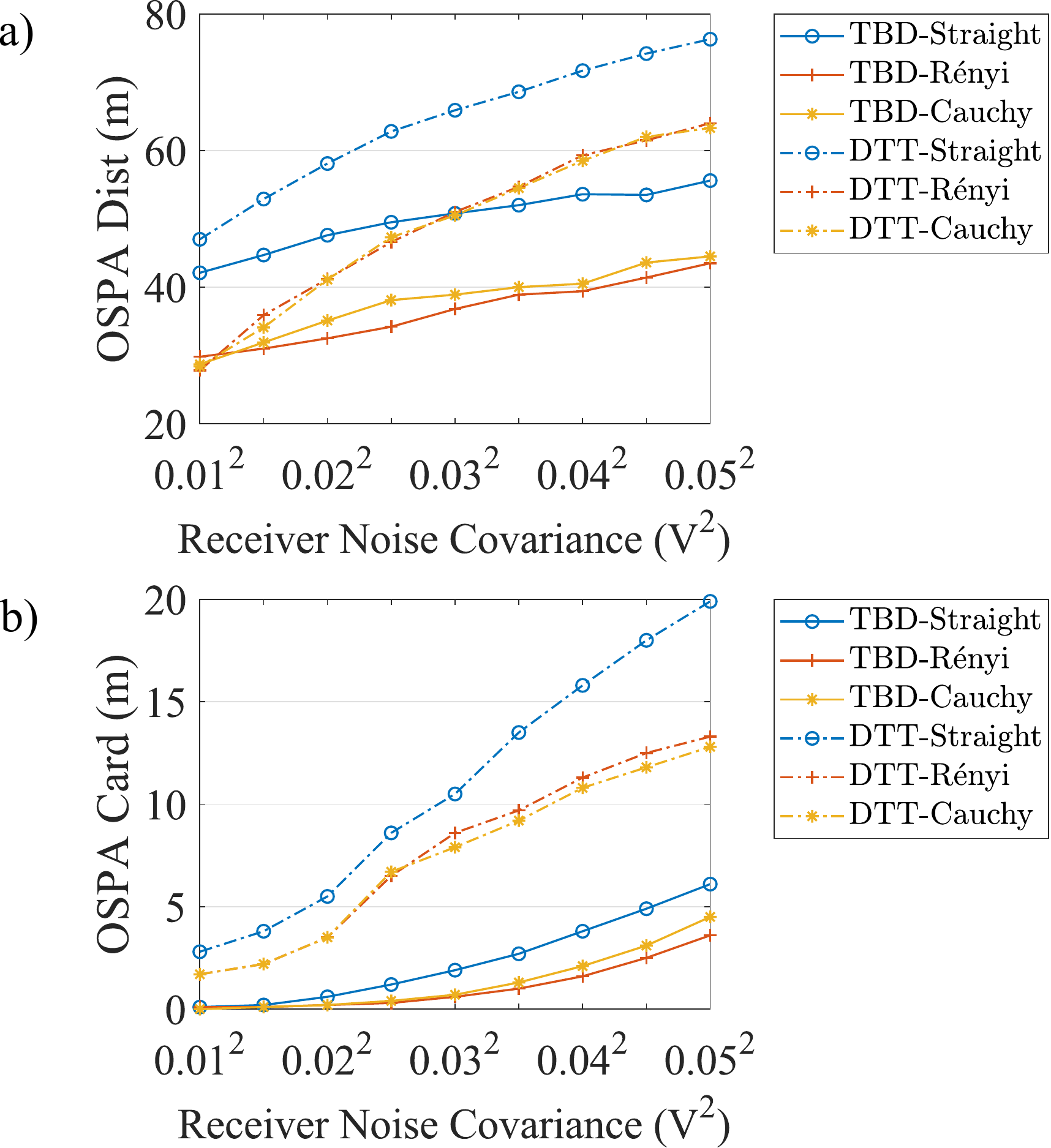}
	\caption{Mean OSPA performance comparison across increasing receiver noise values. 
		Here, -Straight, -R\'{e}nyi and -Cauchy denote straight path, R\'{e}nyi divergence and Cauchy divergence based planning strategies, respectively: a) OSPA distance; b) OSPA cardinality.}
	\label{fig_Ex2_OSPA}
\end{figure}

We use OSPA cardinality and OSPA distance to compare performance across the three  planning strategies for TBD and DTT approaches. We perform 100 Monte Carlo runs for each of the six cases and receiver noise levels $\Sigma_{\eta} =
\{0.010^2, 0.015^2,\dots,0.050^2\}~\text{V}^2$ for the the scenario shown in Fig.~\ref{fig_Ex1_Estimated_Positions}. 
OSPA distance and cardinality results in Fig.~\ref{fig_Ex2_OSPA} show that the proposed path planning for TBD strategy provides significantly better estimation performance over planning for DTT based strategies as demonstrated by the lower OSPA distance metrics in the presence of increasing receiver noise. The TBD approaches are more effective than DTT approaches, especially because  of the failure of DTT methods to detect changes in the number of objects in the presence of birth and death processes as evident in Fig.~\ref{fig_Ex2_OSPA}b.

Intuition suggests that information based approach should execute control actions to continually position the UAV to locations with the best ability to track multiple objects undergoing motion changes. Information based planning strategies outperforming the straight path approaches in both the TBD and DTT methods agrees with this intuition.  Although, R\'{e}nyi or Cauchy divergence as reward functions improve the overall tracking performance compared to the straight path method, we also observe that R\'{e}nyi divergence is more discriminative than Cauchy divergence in our task and yields better OSPA distance values and hence the best performance. 


\section{Conclusion} \label{sec_conclusion}
In this paper, we have proposed an online path planning algorithm for joint detection and tracking of multiple radio-tagged objects under low SNR conditions. The planning for multi-object tracking problem was formulated as a POMDP with two information-based reward functions and the JMS TBD-LMB filter. In particular, the planning formulation incorporates the practical constraint to maintain a safe distance between the UAV and objects of interest to minimize the disturbances from the UAV. We have derived a measurement likelihood for the TBD-LMB filter and proved that the likelihood is separable in practice for multiple radio-tagged objects; thus deriving an accurate multi-object TBD filter. The results demonstrated that our approach is highly effective in reducing the estimation error of multiple-objects in the presence of low signal-to-noise ratios compared to both detection-then-track approaches and tracking without planning.

\appendix[Mathematical Proofs]
The following Lemma facilitates the proof of Proposition~\ref{prop_1}.
\begin{lemma} \label{lemma_iq}
\textit{The STFT of the discrete-time signal $y^{(m)}_k[\cdot]$ can be expressed in terms of in-phase and quadrature forms:
\begin{align}
Y^{(m)}_{k}[l] =\sum_{\mathbf{x}_k \in \mathbf{X}_k} G^{(m,l)}(\mathbf{x}_k) + H^{(m)}_k[l] = Y^{(m)}_{k,I}[l] + j Y^{(m)}_{k,Q}[l] \notag.
\end{align}
Furthermore, the components, $Y^{(m)}_{k,I}[\cdot]$ and $Y^{(m)}_{k,Q}[\cdot]$, are independent non-zero mean Gaussian random variables with covariance $\Sigma_z = E_w \Sigma_{\eta}/2$.
}
\end{lemma}	
\textbf{Proof:}
	 First, we show that the in-phase  and quadrature components of the noise terms $H^{(m)}_k[\cdot]$ of $Y^{(m)}_{k}[\cdot]$  are independent. Next, we prove that the magnitude of the signal term $\sum_{\mathbf{x}_k \in \mathbf{X}_k} G^{(m,\cdot)}(\mathbf{x}_k)$ of $Y^{(m)}_{k}[\cdot]$ has the form $\big|\mu W[\cdot]\big|$ where $\mu$ is zero or a constant and W is as defined in \eqref{eq_fft_win_def}. Therefore, for a given frequency frame $l$, the in-phase and quadrature components of $Y^{(m)}_{k}[\cdot]$ are characterized by constant signal terms of the form $\big|\mu W[\cdot]\big|$ and independent noise terms. Thus, as proven in \cite[pp.17]{richards2007discrete}, the in-phase  and quadrature components are independent since their cross-correlation coefficient is zero. Detailed as below.
	
	Since the receiver noise $\eta\sim\mathcal{N}(\cdot;0, \Sigma_{\eta})$ is narrowband wide-sense-stationary Gaussian, it can rewritten in terms of in-phase and quadrature noise components~\cite[pp.159]{Davenport1987}:
	\begin{align}
		\eta(t) = \eta_I(t) + j \eta_Q(t).
	\end{align}
	where $\eta_I(\cdot)$ and $\eta_Q(\cdot)$ are independent zero mean Gaussian random variables with covariance $\Sigma_{\eta}/2$. Then the STFT transformation of the noise components into time-frequency frames in \eqref{eq_timefreq_noise_frames} follows:
	\begin{align}
		H^{(m)}_k[l] = H^{(m)}_{k,I}[l] + j H^{(m)}_{k,Q}[l],
	\end{align}
	where $H^{(m)}_{k,I}[\cdot]$ and $H^{(m)}_{k,Q}[\cdot]$ are 
	also independent zero-mean Gaussian random variables with covariance $\Sigma_{z} = E_{w}\Sigma_{\eta}/2$, as proven in \cite[pp.10-12]{richards2007discrete}. 
 Thus, by rewriting $Y^{(m)}_{k}[\cdot]$ in \eqref{eq_stft_out_signal_noise} in terms of in-phase and quadrature components, and letting $\Gamma^{(m)}_{k}[l] =\sum_{\mathbf{x}_k \in \mathbf{X}_k} G^{(m,l)}(\mathbf{x}_k)$, for simplicity, we have: 
    \begin{equation}  
	\resizebox{0.5\textwidth}{!}{ $\begin{aligned}
		Y^{(m)}_{k}[l] 
      &= \Gamma^{(m)}_{k}[l] +  H^{(m)}_{k}[l] \\
      &= \big(\text{Re}\{\Gamma^{(m)}_{k}[l]\} + H^{(m)}_{k,I}[l]\big) + j
      \big(\text{Im}\{\Gamma^{(m)}_{k}[l]\} + H^{(m)}_{k,Q}[l]\big)   \\
      &= Y^{(m)}_{k,I}[l] + j Y^{(m)}_{k,Q}[l],\end{aligned} $.}
	\end{equation}
    
	From the initial assumption in Proposition \ref{prop_1}, $C(\mathbf{x}_k) \cap C(\mathbf{x}'_k) = \emptyset ~ \forall~\mathbf{x}_k,\mathbf{x}'_k \in \mathbf{X}_k $. Thus, $(m,l) \notin C(\mathbf{x}_k)\cap C(\mathbf{x}'_k)$. In other words, at time-frequency frame $(m,l)$, at most one
	object $\mathbf{x}_k \in \mathbf{X}_k$ contributes to the magnitude of $|\Gamma^{(m)}_k[l]|$, such that:	 
	\begin{equation}  \label{eq_total_received_magnitude}
	\resizebox{0.5\textwidth}{!}{ $\begin{aligned}
		|\Gamma^{(m)}_{k}[l]| &= |\sum_{\mathbf{x}_k \in \mathbf{X}_k} G^{(m,l)}(\mathbf{x}_k)| =%
		\begin{cases}
		|G^{(m,l)}(\mathbf{x}_k)| & (m,l) \in C(\mathbf{x}_k) \\ 
		0 & \text{otherwise,}%
		\end{cases}  \end{aligned} $ }
	\end{equation}
	where, following the signal model illustrated in Fig.~\ref{fig_received_pulse_stft_illustration},
	\begin{equation}  \label{eq_intensity_abs}
	\resizebox{0.5\textwidth}{!}{ $\begin{aligned}
		|G^{(m,l)}(\mathbf{x}_k)| &= 
		\begin{cases}
		\big|\gamma(\zeta_k)W[l-l^{(\lambda_k)}]\big| & \text{if } m \in \{m^{(\tau_k)},m^{(\tau_k)}+1\} \\ 
		0 & \text{otherwise,}
		\end{cases}\\
		C(\mathbf{x}_k)&=\{m^{(\tau_k)},m^{(\tau_k)}+1\} \times S(l^{(\lambda_k)}), \end{aligned} $ }
	\end{equation}
	and $S(l^{(\lambda_k)})\subseteq \{0,\dots,L-1\} $ denotes the window function---see Table~\ref{table_L_scale_window_type}---dependent number of frequency samples contributed by object $\mathbf{x}_k$ . 
	
According to \eqref{eq_total_received_magnitude} and \eqref{eq_intensity_abs}, $|\Gamma^{(m)}_{k}[\cdot]|$ is deterministic and has the form $\big|\mu W[\cdot]\big|$, where $\mu$ is zero or a constant. Consequently, the cross-correlation coefficient $\rho_{IQ}$ of $Y^{(m)}_{k,I}[\cdot]$ and $Y^{(m)}_{k,Q}[\cdot]$ is zero, as proven in \cite{richards2007discrete}:
	\begin{align}
		\rho_{IQ} = \Big(\mathbb{E}\big(Y^{(m)}_{k,I}[l]Y^{(m)}_{k,Q}[l]\big) - \Gamma^{(m)}_{k,I}[l] \Gamma^{(m)}_{k,Q}[l] \Big)/\Sigma_{z} = 0
	\end{align}
	Therefore, $Y^{(m)}_{k,I}[\cdot]$ and $Y^{(m)}_{k,Q}[\cdot]$ are both independent
	non-zero mean Gaussians with the same covariance $\Sigma_z = E_{w}\Sigma_{\eta}/2$. \scalebox{1.5}{$\Box$} 
	
\textbf{Proof of Proposition \ref{prop_1}: }
	Applying Lemma \ref{lemma_iq}, for any time-frequency frame $(m,l)$, $Y^{(m)}_{k,I}[l]$ and $Y^{(m)}_{k,Q}[l]$ are independent non-zero mean Gaussian. Thus, combining the result in \cite[pp.17-18]{richards2007discrete}, if object $\mathbf{x}_k$ contributes to the measurement $z_k$ at time-frequency frame $(m,l)$: $|\Gamma^{(m)}_{k}[l]| = |G^{(m,l)}(\mathbf{x}_k)|$, then the measurement likelihood function of $z^{(m,l)}_{k} = |Y^{(m)}_{k}[l]|$ is: 
	\begin{align}
	p(z^{(m,l)}_{k}|\mathbf{x}_k) &= \varphi(z^{(m,l)}_k;|G^{(m,l)}(\mathbf{x}_k)|,\Sigma_z),
	\end{align}
	where $
	\varphi(x;\nu,\Sigma) = {x} \exp \{-({x^2 + \nu^2})/({2\Sigma%
	})\} I_0({x\nu}/{\Sigma})/{\Sigma}
	$
	is a \textit{Ricean} distribution; $I_0(\cdot)$ is the Bessel
	function of the first kind defined as $I_0(x) = \sum\limits_{j=0}^{\infty} (-1)^j(x^2/4)^j/{(j!)^2}.$

	When no signal contributes to a frame $(m,l)$, $|\Gamma^{(m)}_{k}[l]| = 0$; then the measurement likelihood function of $z^{(m,l)}_{k}$ is: 
	\begin{align}  \label{eq_mag_zero_dist}
	p(z^{(m,l)}_{k}|\mathbf{x}_{k}) &= \phi(z^{(m,l)}_{k};\Sigma_z),
	\end{align}
	where 
	$\phi(x;\Sigma) = {x} \exp \{-{x^2}/{\Sigma}\}/{\Sigma}$
	is a \textit{Rayleigh} distribution.	
	
	Thus, at any given frame $(m,l) \in \{0,\dots,M-1\} \times \{0,\dots,L-1\}$, the
	measurement likelihood function of $z^{(m,l)}_{k} = |Y^{(m)}_k[l]|$%
	, given object state $\mathbf{x}_{k}$ follows: 
	\begin{equation}  \label{eq_single_likelihood}
	\resizebox{0.5\textwidth}{!}{ $\begin{aligned} \notag
		p(z^{(m,l)}_k|\mathbf{x}_k )=
		&\begin{cases} \varphi(z^{(m,l)}_k; |G^{(m,l)}(\mathbf{x}_k)|,\Sigma_z) & (m,l) \in C(\mathbf{x}_k), \\
		\phi(z^{(m,l)}_k;\Sigma_z) & (m,l) \notin
		C(\mathbf{x}_k).\\ \end{cases}  \end{aligned} $ }
	\end{equation}
	
	Since there is no overlap between the influence regions of two objects, \ie, $C(\mathbf{x}_k) \cap C(\mathbf{x}'_k) = \emptyset ~ \forall~\mathbf{x}_k,\mathbf{x}'_k \in \mathbf{X}_k $,  the measurement likelihood of $z_k$ conditioned on the multi-object state $\mathbf{X}_{k}$, can be modeled as a separable function:
	\begin{align}
	&g(z_{k}|\mathbf{X}_{k}) 
	=  \Big(\prod\limits_{\mathbf{x}_k \in \mathbf{X}_k} \prod\limits_{(m,l) \in C(\mathbf{x}_k)  }\varphi (	z_k^{(m,l)};|G^{(m,l)}(\mathbf{x}_k)|,\Sigma_{z})\Big) \notag\\ &~~~~~~~~~~~~~\times \prod\limits_{(m,l) \notin \cup_{\mathbf{x}_k \in \mathbf{X}_k} C(\mathbf{x}_k)  } \phi (z_k^{(m,l)};\Sigma_{z})  \\
	&= \prod\limits_{(m,l)=(0,0)}^{(M-1,L-1)} \phi (z_k^{(m,l)};\Sigma_{z}) \prod\limits_{\mathbf{x}_k \in \mathbf{X}_k}g_{{z_k}}(\mathbf{x}_k)\propto
	\prod\limits_{\mathbf{x}_k \in \mathbf{X}_k}g_{{z_k}}(\mathbf{x}_k),\notag
	\end{align}
	where 
	\begin{equation} \notag
	g_{z_k}(\mathbf{x}_k)=\prod\limits_{(m,l)\in C(\mathbf{x}_k)}\dfrac{\varphi (%
		z_k^{(m,l)};|G^{(m,l)}(\mathbf{x}_k)|,\Sigma_{z})}{\phi (%
		z_k^{(m,l)};\Sigma_{z})}. \scalebox{1.5}{$\Box$}
	\end{equation}

\ifCLASSOPTIONcaptionsoff
\newpage \fi

\end{document}